\documentclass[preprint,12pt,authoryear,times]{elsarticle}

\usepackage[T1]{fontenc}
\usepackage{floatrow}
\usepackage[greek,english]{babel}

\usepackage{epigraph}
\usepackage{graphicx}
\graphicspath{{.}}
\usepackage{comment}
\usepackage[utf8]{inputenc}
\usepackage[T1]{fontenc}
\usepackage{xcolor}
\usepackage[hidelinks,colorlinks=true]{hyperref}
\usepackage{fullpage}
\usepackage[right]{lineno}
\usepackage{amssymb}
\usepackage{mathtools}
\usepackage{amsmath} 
\usepackage{graphicx}
\graphicspath{{./images}}
\usepackage[mediumspace,mediumqspace,squaren]{SIunits}

\usepackage[capitalise,nameinlink,sort&compress]{cleveref}
\crefname{equation}{}{}
\crefname{appendix}{}{}
\creflabelformat{equation}{#2#1#3}
\crefmultiformat{equation}{#2#1#3}{, #2#1#3}{, #2#1#3}{, #2#1#3}
\renewcommand\eqref[1]{(\cref{#1})}

\DeclareMathOperator{\Div}{Div}
\DeclareMathOperator{\Grad}{Grad}
\DeclareMathOperator{\grad}{grad}
\DeclareMathOperator{\divergence}{div}
\DeclareMathOperator{\Tr}{tr}

\DeclareMathOperator{\diag}{diag}

\newcommand{\matr}[1]{\mathbf{#1}}
\newcommand{\E}{{\text{e}}}
\renewcommand{\vec}[1]{\mathbf{#1}}
\newcommand{\diff}[1]{\mathrm{d}{#1}}
\newcommand{\todo}[1]{{\color{blue}{[TODO: \textit{#1}]}}}

\newcommand{\pp}[1]{\left({#1}\right)}
\newcommand{\bb}[1]{\left[ #1 \right]}
\newcommand{\eqdef}{\vcentcolon =}
\newcommand{\pdiff}[2]{\frac{\partial{#1}}{\partial{#2}}}
\newcommand{\ppdiff}[2]{\frac{\partial^2{#1}}{\partial{#2}^2}}

\newcommand{\linepdiff}[2]{{\partial{#1}}/{\partial{#2}}}

\newcommand{\linefdiff}[2]{{\diff{#1}}/{\diff{#2}}}
\newcommand{\fdiff}[2]{\frac{\diff{#1}}{\diff{#2}}}
\newcommand{\ramp}[1]{\pp{#1}_+}

\newcommand\restr[2]{{
  \left.\kern-\nulldelimiterspace
  #1 
  \vphantom{\big|} 
  \right|_{#2} 
  }}

\usepackage{amssymb}

\journal{Journal of the Mechanics and Physics of Solids}

\begin{document}

\begin{frontmatter}



\title{\textbf{Hydromechanical field theory of plant morphogenesis}}


\author[mpipz,mpicbg,csbd]{Hadrien Oliveri\corref{cor1}}

\author[inst2,rdp]{Ibrahim Cheddadi\corref{cor2}}

 \cortext[cor1]{ \href{mailto:holiveri@mpipz.mpg.de}{holiveri@mpipz.mpg.de}}

\cortext[cor2]{\href{mailto:ibrahim.cheddadi@univ-grenoble-alpes.fr}{ibrahim.cheddadi@univ-grenoble-alpes.fr}}

\affiliation[mpipz]{organization={Max Planck Institute for Plant Breeding Research},
            city={Cologne},
            postcode={50829}, 
            country={Germany}}

\affiliation[mpicbg]{organization={Max Planck Institute of Molecular Cell Biology and Genetics},
            city={Dresden},
            postcode={01307}, 
            country={Germany}}

\affiliation[csbd]{organization={Center for Systems Biology Dresden},
            city={Dresden},
            postcode={01307}, 
            country={Germany}}
            
\affiliation[inst2]{organization={Université Grenoble Alpes, CNRS, Grenoble INP, TIMC},
            city={Grenoble},
            postcode={38000}, 
            country={France}}

\affiliation[rdp]{organization={Laboratoire de Reproduction et Développement des Plantes, Université de Lyon, ENS de Lyon, UCBL, INRAE, CNRS, Inria},
            city={Lyon},
            postcode={69364}, 
            country={France}}

\begin{abstract}The growth of plants is a hydromechanical phenomenon in which cells enlarge by absorbing water, while their walls expand and remodel under turgor-induced tension. In multicellular tissues, where cells are mechanically interconnected, morphogenesis results from the combined effect of local cell growths, which reflects the action of heterogeneous mechanical, physical, and chemical fields, each exerting varying degrees of nonlocal influence within the tissue. To describe this process, we propose a  physical field theory of plant growth. This theory treats the tissue as a \textit{poromorphoelastic} body, namely a growing poroelastic medium, where growth arises from pressure-induced deformations and osmotically-driven imbibition of the tissue. From this perspective, growing regions correspond to hydraulic sinks, leading to the possibility of complex non-local regulations, such as water competition and growth-induced water potential gradients. More in general, this work aims to establish  foundations for a mechanistic, mechanical field theory of morphogenesis in plants, where growth arises from the interplay of multiple physical fields, and where biochemical regulations are integrated through specific physical parameters.
\end{abstract}
\begin{keyword}
plant mechanics \sep morphogenesis \sep  growth  \sep elasticity \sep morphoelasticity \sep poroelasticity
\MSC[2020] 74B20 \sep 74F10 \sep 74F20 \sep 92B99
\end{keyword}

\end{frontmatter}






\epigraph{\textit{Einfach schlief in dem Samen die Kraft; ein beginnendes Vorbild\\ Lag, verschlossen in sich, unter die Hülle gebeugt,\\
Blatt und Wurzel und Keim, nur halb geformet und farblos; \\ Trocken erhält so der Kern ruhiges Leben bewahrt, \\ Quillet strebend empor, sich milder Feuchte vertrauend,\\
Und erhebt sich sogleich aus der umgebenden Nacht.}}{Johann Wolfgang von Goethe, \textit{Die Metamorphose der Pflanzen}, 1798}

\section{Introduction}

\subsection{Modelling the growth of plant tissues}
Morphogenesis is the fundamental process by which biological organisms establish their shape. This phenomenon relies upon multiple genetic, chemical, mechanical, and physical regulations, which are typically multiscale, thermodynamically open, out-of-equilibrium, and coupled in a so-called complex system. 
To achieve a rational understanding of morphogenesis, dedicated models are indispensable; in particular, a grand challenge is to build mechanistic continuum theories that describe the emergent physical behaviour of living tissues at the macroscopic scale.

In plants, the key phenomenon driving morphogenesis is \textit{growth}, the general process by which a physical body deforms by gaining mass \citep{Goriely2017}. Indeed, in plant tissues, adjoining cells are mechanically interconnected and constrained by an extracellular cell wall matrix which prevents cell migration and topological rearrangements \citep{hamant2010mechanics,boudaoud2010introduction,coen2023mechanics}. Therefore, complex organ shapes result from the anisotropic and heterogeneous growth (and division) of the cells, generating large solid deformations of connected tissue regions \citep{coen_genetics_2004,boudaoud2010introduction}.

The theory of \textit{morphoelasticity} is a modern mechanical theory of growth, which has emerged historically from the nonlinear field theories of elasticity and elastoplasticity  \citep{Goriely2017}. 
This framework has been adapted with remarkable success to a wealth of biological systems \citep[see the reviews by][]{menzel2012frontiers,kuhl2014growing,ambrosi2019growth}, and provides a natural framework to model plant growth at the tissue scale \citep{vandiver2008tissue,dervaux2008morphogenesis,goriely2010elastic,holland2013mechanics,moulton2020multiscale,boudaoud2023multiscale,Oliveri2024,liang2009shape}. Historically, an important problem in growth theory has been the characterisation of growth-induced mechanical instabilities, which refer to the loss of stability and  change in configuration undergone by a growing tissue as a result to growth-induced internal stresses \citep[e.g.][]{BenAmar2005,goriely1998spontaneous,liang2009shape,dervaux2008morphogenesis,wang2015three,liu2013pattern,jia2018curvature}. In this context, the growth deformation is treated typically as a constant bifurcation parameter controlling the onset of instability. In this respect, the focus lies on the static  elasticity problems arising from growth, rather than the physical origin and kinetics of growth itself. 
In contrast, the so-called problem of \textit{morphodynamics} consists in studying growth as a variable of a dynamical system whose evolution reflects the physical state of the tissue, reflecting the inherently dynamic nature of developmental processes  \citep{vandiver2009morpho,chickarmane2010computational}.

From a biological point of view, plant growth is controlled dynamically through a multitude of genes and hormones that form heterogeneous patterning fields \citep{coen2000art,johnson2011genetic}. To address the role of genetic patterning in morphogenesis, continuum computational models have been proposed, based on the so-called notion of \textit{specified growth}, detailed in \cite{coen_genetics_2004,kennaway_generation_2011,kennaway2019volumetric} and applied in various context, see 
\cite{kierzkowski2019growth,green_genetic_2010,whitewoods2017growth,Rebocho2017,lee2019shaping,zhang2020wox,whitewoods2020evolution,peng2022differential,KellyBellow2023,zhang2024mechanism} \citep[see also][for reviews]{ali2014physical,coen2017genes,mosca2018modeling,coen2023mechanics}. Specified growth refers to the \textit{intrinsic} growth a volume element undergoes when isolated from the rest of the tissue, reflecting some local biological identity (e.g. gene expression, polarity or, more abstractly, the action of morphogens) which control the kinematics of growth. Then, given a specified growth field, a compatible deformation can be obtained in a second step, by minimising the body's elastic energy. 
This approach has been instrumental in demonstrating how spatially heterogeneous gene expression patterns control the emergence of complex shapes in three-dimensions. However, from a biophysical perspective, specified growth remains a fundamentally phenomenological and essentially kinematic representation of growth, as its precise link to cellular mechanics--specifically the explicit physical action of genes--is not fully characterised.

In contrast, the aim of biophysics is to explain growth from physical and mechanical principles. In this context, physically-based discrete models  describe tissues at the cellular level, incorporating detailed cellular structure and geometry, mechanical anisotropies, and fine mechanisms underlying growth   \citep{khadka2019feedback,rudge2005computational,dupuy2007system,hamant2008developmental,corson_turning_2009,fozard_vertex-element_2013,boudon_computational_2015,cheddadi2019coupling,daher2018anisotropic,zhao2020microtubule,long2020cellular,sassi_auxin-mediated_2014,fridman2021root,LIU20221974,peng2022differential,KellyBellow2023,ali2019simulating,merks2011virtualleaf,alim2012regulatory,bessonov2013deformable,bassel2014mechanical}. The goal in these models is then to reproduce the growth phenomenon from first principles, thereby allowing for rigorous testing of hypotheses at the cellular scale through  specific biophysical parameters.  However, this type of approach relies on a discretised representation of the tissue, thus it is often doomed to remain computational. While cellular models offer detailed insights, they do not enjoy the generality, minimalism, scalability, and analytic tractability of a continuum mathematical theory.

\subsection{The hydromechanical basis of cell growth}

To build such a theory on physical grounds, we start with the basic physics of cell expansion. Plant cells grow by absorbing water from their surrounding—a process driven by their relatively high osmolarity \citep{schopfer2006biomechanics,geitmann_mechanics_2009,ortega2010plant}. This phenomenon is captured by the historical model of \cite{lockhart1965analysis}, later expanded in a somewhat similar manner by \cite{cosgrove1985cell,cosgrove1981} and \cite{ortega_augmented_1985} \cite[see also the discussions by][]{goriely_elastic_2008,dumais2021,forterre2022basic,forterre2013slow,ali2023revisiting,cheddadi2019coupling,ortega2010plant,geitmann_mechanics_2009}. This model describes the elongation of a single cylindrical cell of length $\ell$. The expansion rate $\dot \ell $ of such cell can be expressed in terms of the volumetric influx of water through \citep{DAINTY1963279}
\begin{equation}
\dot \ell   = k_c \ell \pp{\pi - p} , \label{eqn:ortega-fluxes}
\end{equation}
with $k_c$  the effective hydraulic conductivity the cell;  $\pi$ and $p$ respectively the excess osmotic and hydrostatic pressures with respect to the outside; and where the overdot denotes differentiation w.r.t. time $t$. The quantity $\psi = p-\pi $, the \textit{water potential} of the cell relative to the outside, characterises the capacity of the cell to attract water. Note that \eqref{eqn:ortega-fluxes} does not define a closed system, as the pressure $p$ is related to the mechanics of the cell wall via the balance of forces. The growth of plant cells is a complex process that involves loosening, yielding under tension, and remodelling of the cell walls \citep{cosgrove2005growth,cosgrove2016plant,cosgrove2018diffuse}. This process is  modelled as a Maxwell-type viscoelastoplastic creep. In an elongating cylindrical cell, stress can be explicitly written in terms of pressure, and this rheological law can be  written as
\begin{equation}
   \frac{\dot \ell }{\ell}  = \Phi \ramp{p - y} + \frac{\dot p}{E} ,\label{eqn:ortega}
\end{equation}
with $\ramp{x}\eqdef\max\pp{x,\cdot}$  the ramp function; $y$ a threshold pressure above which growth occurs; $\Phi$ the so-called \textit{extensibility} of the cell; and $E$ the effective Young's modulus of the cell. Equations \eqref{eqn:ortega,eqn:ortega-fluxes} now form a closed system for $\ell$ and $p$, illustrating that the turgor pressure is not a control parameter of the rate of growth, but a dependent variable of the cell expansion process. Indeed, in the steady growth regime ($\dot p = 0$), the pressure is fully determined by the three parameters $y$, $\pi$ and $\alpha_c \eqdef  k_c / (k_c + \Phi)$, and can be eliminated from the system:
\begin{equation}\label{eqn:lockhart-pressure}
    \frac{\dot \ell}{\ell} = \alpha_c\Phi \pp{\pi - y}, \quad 
    p = \alpha_c\pi +  (1-\alpha_c)y
\end{equation}
(with $y\leq p \leq \pi$).

Recently, the Lockhart model served as a starting point for a vertex-based, multicellular hydromechanical model of plant morphogenesis, including water movements between cells, and mechanically-driven cell wall expansion 
\citep{cheddadi2019coupling}. This generalisation of the single-cell model generates several additional emergent spatiotemporal effects that arise from the coupling between wall expansion and water movements within a large developing tissue, such as hydraulic competition effects and water gradients. 
%
%
Overall, the relationship between water and growth, as highlighted by the Lockhart-Ortega-Cosgrove model and its multidimensional generalisation, is central to the physical functioning of plant living matter, and the basic notion that water must travel from sources (e.g. vasculature) to growing regions is a fundamental principle  of development.

\subsection{Hydromechanical field theory for plant growth}
Building on this notion, we here develop a physical field theory of plant morphogenesis including tissue mechanics, wall synthesis, and hydraulics. Viewing the tissue as a network of cells exchanging water, a natural formalism is the theory of porous media, which describes the deformations of fluid-saturated materials. The modern theory of porous media has a complex history originating in the seminal works of Fick and Darcy on diffusion, and Fillunger, Terzaghi and Biot on soil consolidation, formalised later within the context of the theory of mixtures, notably based on contributions by Truesdell and Bowen  \citep{de1992development,de2012theory,bedford1983theories}. There is now a rich body of literature exploring the application of these concepts to biology in various contexts, including biological growth. 
However, despite the clear significance of water in plant mechanics \citep{forterre2022basic,dumais2012vegetable}, the role of hydromechanics in the context of plant growth modelling has but marginally been examined. Several notable  works have featured linear poroelastic descriptions of plant matter, occasionally including growth \citep{molz1974water,molz1975dynamics,molz1978growth,plant1982continuum,passioura2003tissue,philip1958propagation}; however, these studies are limited to simple geometries and have not culminated in a general three-dimensional theory of growth. While the limiting role of water fluxes within plant tissues has long been debated \citep{boyer1988cell, cosgrove1993water} and remains challenging to assess experimentally, recent years have seen a resurgence of interest in this problem, with new studies supporting the potential for hydromechanical control in tissue development \citep{laplaud2024assessing, alonso2024water, long2020cellular,cheddadi2019coupling}. 
Building on these recent advancements, we propose a hydromechanical model for plant tissues seen as morphoelastic, porous materials encompassing the solid matrix of the cells and their fluid content. 
Technically speaking, our approach is analogous to previous models, for instance for tumour growth \citep{XUE2016409,fraldi2018cells} or brain oedema \citep{lang2015propagation}, however the specificities of plant living matter demand a dedicated treatment. Overall, this work aims to bridge the conceptual gap between cell biophysics and growth at the tissue scale, by describing growth as a consequence of simultaneous mechanical deformations and imbibition of the cells, thereby extending the seminal analysis of Lockhart to the framework of continuum media.  

The goals are then (i) to establish theoretical foundations for a generic theory of plant growth within the framework of nonlinear continuum mechanics (\cref{general}); and (ii) to characterise the guiding principles that emerge from multiple couplings between mechanical, physical and hydraulic fields (\cref{rod,tube,seed}).





\section{General theory\label{general}}



\subsection{Geometry and kinematics\label{geometry}}


\subsubsection{Total deformation}
We describe the geometry and kinematics of the evolving tissue viewed as a continuum domain embedded in the three-dimensional space \citep[see][for details on finite strain theory]{holzapfel2000nonlinear,Goriely2017}. We introduce the initial configuration $\mathcal B_0 \subset \mathbb R^3$ at time $t=0$ of the body, and its current configuration $\mathcal B\subset \mathbb R^3$ at $t\geq 0$; see \cref{fig:fig1}(a). The deformation of the body from $\mathcal B_0$ to $\mathcal B$ is described by a smooth map $\boldsymbol\chi: \mathcal B_0 \times \mathbb R \rightarrow \mathcal B$ that sends {material} points $\vec X\in \mathcal B_0$ onto {spatial} points $\vec x = \boldsymbol{\chi}(\vec X,t)\in \mathcal B$, at a given time $t\in \mathbb R$. 
The {deformation gradient tensor} is defined as $\matr F \eqdef \Grad \boldsymbol\chi$,
with $\Grad $ the {material} (or {Lagrangian}) gradient, with respect to the reference configuration. 
Finally, the Jacobian determinant of the deformation, $J \eqdef \det \matr F$, measures the local volumetric expansion due to $\boldsymbol\chi$ (i.e. $\diff v = J \diff V$). 


Next, we describe the kinematics of the deformation by introducing the material and spatial (or Eulerian) velocities, respectively $\vec V$ and $\vec v$, defined as $ \vec V (\vec X, t)=\vec v (\vec x, t)= \linepdiff{\boldsymbol \chi (\vec X, t)}{t}$.
A natural measure of the relative rate of deformation is then the gradient of spatial velocity,
\begin{equation}
    \matr L  \eqdef \grad \vec v = \matr {\dot F}\matr F^{-1},\label{eqn:velocity-gradient}
\end{equation}
with $\grad (\cdot) = \matr F^{-\top}\Grad (\cdot)$  the Eulerian gradient with respect to the current configuration. 
We also recall the standard kinematic formulae
$
  {\dot J} / J= \divergence \vec v =
  \Tr \matr L$, 
with $\divergence$ the Eulerian divergence and $\Tr$ the trace operator. 
Here, the overdot denotes the material derivative $\dot J = \linefdiff{J}{t}$, related to the Eulerian derivative $\linepdiff{J}{t}$ through the generic formula
\begin{equation}
    \dot J = \pdiff{J}{t} + \vec v \cdot \grad J. \label{eqn:material-derivative}
\end{equation}

\subsubsection{Growth}

\begin{figure*}[ht!]
    \centering
\includegraphics[width=\linewidth]{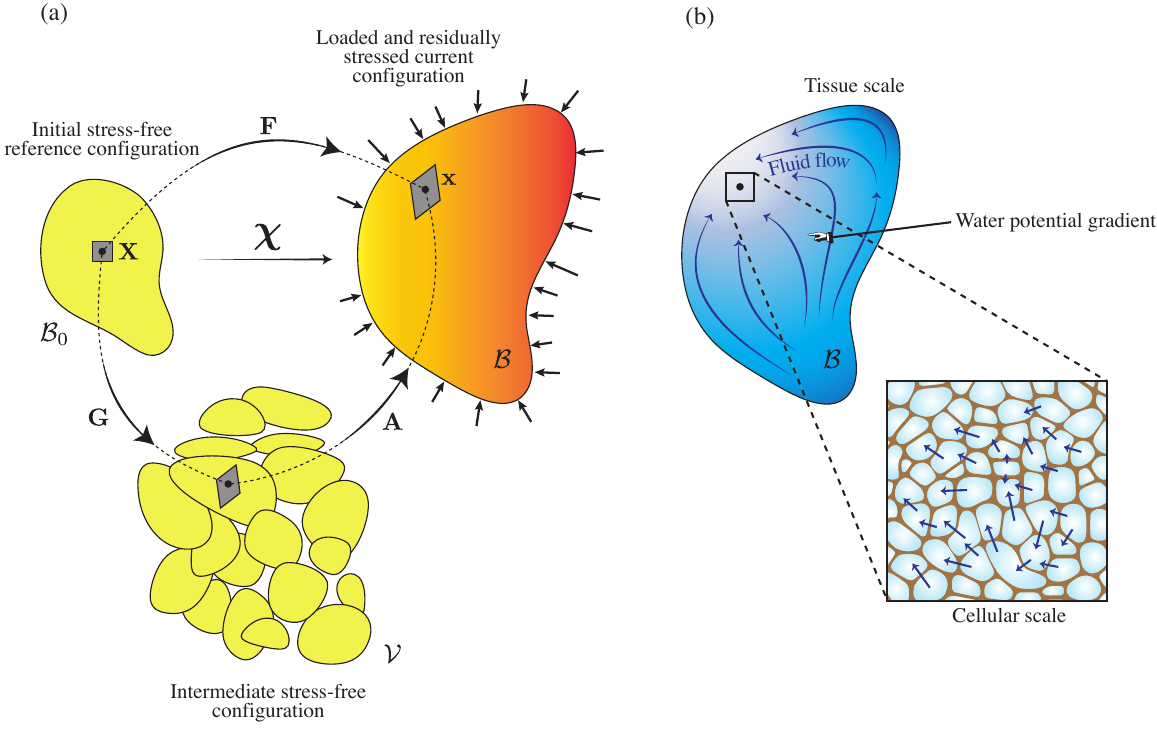}
\caption{Poromorphoelastic theory of plant morphogenesis. (a) The multiplicative decomposition of morphoelasticity \cite[redrawn after][]{Goriely2017}. Starting from a stress-free initial configuration, a local growth deformation $\matr G$ is applied on volume elements, resulting in an incompatible intermediate  configuration. A second deformation $\matr A$ allows to ensure compatibility, and results in a stressed configuration that is loaded externally and includes residual growth stresses. (b) The material is modelled as a growing porous medium, capturing the flow of water across the domain.}
    \label{fig:fig1}
\end{figure*}

We model growth using the framework of morphoelasticity \citep{Goriely2017}. The fundamental postulate of morphoelasticity is the multiplicative decomposition of deformations into an elastic deformation tensor $\matr A$ and a growth tensor $\matr G$   \citep{rodriguez1994stress}: 
\begin{equation}
    \matr F = \matr A  \matr G ;\label{eqn:multiplicative}
\end{equation} 
see \cref{fig:fig1}(a).
Constitutively, the growth deformation $\matr G$ is assumed to be \textit{anelastic}, meaning it does not contribute to the strain energy, and describes the configurational change in the local reference configuration of the body due to local mass addition. In contrast, the elastic deformation $\matr A$ generates stresses and enters the strain energy function. Note that there is no requirement for the growth deformation to be compatible, that is, in general there is no deformation embeddable in the Euclidean space of which $\mathbf G$ is the gradient, unlike $\matr F$ which is compatible by definition \cite[see][Chapter 12]{Goriely2017}.  
We define the Jacobian determinants $J_A \eqdef \det \matr A$ and $J_G \eqdef \det \matr G$, so that \eqref{eqn:multiplicative}
\begin{equation}
    J = J_AJ_G;\label{eqn:productdet}
\end{equation} 
and the rate of growth $\matr L_G \eqdef \matr{\dot G}\matr G ^{-1}$ and rate of elastic deformation $\matr L_A \eqdef \matr{\dot A}\matr A ^{-1}$, with
\begin{equation}
    \matr L = \matr L_A +   \matr A \matr L_G \matr A^{-1} .\label{eqn:rates}
\end{equation}
Finally, we introduce $\Gamma \eqdef \Tr {\matr L_G} = \dot J_G/J_G$, measuring the relative rate of volumetric expansion due to growth.

Note that, in the context of plant morphogenesis, most authors have restricted their attention to linear elasticity. In contrast, as we do not require $\matr A$ or $\matr G$ to be small here, our approach is geometrically exact and can capture arbitrarily large strains, and rich elastic behaviours such as strain-stiffening effects \citep{kierzkowski_elastic_2012}.

\subsection{Balance laws}

\subsubsection{Balance of mass\label{balance-mass}}

We model the growing tissue as a \textit{poromorphoelastic} solid representing the cell matrices and the water content; see \cref{fig:fig1}(b). Specifically, the tissue is viewed as a triphasic porous medium composed of a solid phase representing the extracellular cell wall matrix (`$s$'), a pure fluid phase (`$f$'), and the cytoplasmic osmolytes (`$o$'). We define $\vec v_s$, $\vec v_f$, and $\vec v_o$, the Eulerian velocities of the respective phases, with the solid velocity chosen as the reference:
\begin{equation}
    \vec v_s = \vec v. \label{eqn:v-vs}
\end{equation}
We denote with $\phi_s$, $\phi_f$, $\phi_o$ the volume fractions of the solid, fluid and osmolyte components, respectively, in the current configuration (that is, the true volume occupied by each component per unit volume of  tissue). Assuming that the material is saturated, we have
\begin{equation}
    \phi_s + \phi_f + \phi_o = 1. \label{eqn:saturation}
\end{equation}
The pore space volume occupancy is measured by the Eulerian porosity $\phi \eqdef 1 - \phi_s $, which represents the ratio of pore space volume to current tissue volume. Similarly, the intermediate porosity $\Phi \eqdef J_A \phi $ measures the \textit{current} pore space volume per unit \textit{intermediate} volume. The Lagrangian porosity $\Phi^0 \eqdef J\phi= J_G \Phi$ measures the current pore space volume per unit reference volume. 

Here, we assume that all phases are \textit{inherently} elastically incompressible \citep{lang2015propagation}. This assumption implies that in a  {macroscopic} elastic deformation, the pores will shrink or dilate so as to conserve the volume of the solid phase. Thus, we can relate the solid volume fraction in the unstressed configuration ($\Phi_s  $, the solid volume fraction observed upon relieving the stresses) to that in the current configuration $\phi_s$, through
\begin{equation}
    \Phi_s = J_A\phi_s = J_A - \Phi .\label{eqn:phisdeformation}
\end{equation}
During growth and remodelling of the tissue, cells secrete new solid material, divide and form new cell walls that may affect the structure of the cell matrix locally,  thus, the porosity. To represent this effect, we assume generally that the solid volume fraction may evolve as
\begin{equation}
  \dot \Phi_s  = \mathcal H(\vec x, t, \Phi_s, \matr G,\dots),\label{eqn:masss}
\end{equation}
where $\mathcal H$ is a function of the different variables that symbolises the local evolution of the matrix structure during growth. For example, in a scenario where cells do not divide but maintain a constant cell wall thickness, we can assume $\mathcal H = - \Gamma \Phi_s/3$. We assume here that the tissue maintains locally a constant homeostatic cellular structure, so that $\mathcal H=0$ and $\Phi_s$ is a constant. 
We can rewrite \eqref{eqn:masss} using \eqref{eqn:productdet,eqn:material-derivative,eqn:phisdeformation} as
\begin{equation}
   \pdiff{\phi_s}{t}+ \divergence( \phi_s \vec v_s)=  \mathcal H/J_G + \Gamma \phi_s := \xi_s, \label{eqn:masss2}
\end{equation}
with $\xi_s$ representing the source of solid material. 
 Here we assume that the availability of the solid material required for growth is non-limiting, i.e., we ignore the physiological aspects of biosynthesis underlying growth (photosynthesis, nutrient uptake).  

Similarly, the balance of fluid mass is expressed through
\begin{equation}
    \pdiff{\phi_f}{t} +   \divergence (\phi_f\vec v_f) = \xi_f, \label{eqn:massf}
\end{equation} 
where $\xi_f$ is a possible source of water that accounts, typically, for a bulk vascularisation of the tissue. A possible constitutive law for $\xi_f$ is discussed later in this paper. 
Lastly, the balance of mass for the osmolytes is expressed through
\begin{equation} \label{eqn:balance-osmolyte-mass}
    \pdiff{\phi_o}{t} + \divergence\pp{\phi_o\vec v_o}  =  \xi_o.
\end{equation}
with $\xi_o$ representing the synthesis/uptake of osmolytes. Summing \eqref{eqn:balance-osmolyte-mass,eqn:massf,eqn:masss2}  using \eqref{eqn:saturation,eqn:v-vs}, we obtain a single volume balance equality  \citep{lang2015propagation}:
\begin{equation}\label{eqn:sum-balance-mass}
    \divergence\pp{\phi_s\vec v + \phi_f\vec v_s + \phi_o\vec v_o} = \xi_s + \xi_f + \xi_o.
\end{equation}

\subsubsection{Balance of momentum}
Since the growth and fluid transport happen at timescales (hours) that are much longer than the timescale of elastic relaxation (seconds), we  neglect inertia. 
Thus, we write the balance of momentum in terms of the Cauchy stress tensor $\matr T$ as
\begin{equation}
    \divergence \matr T + \vec b = \vec 0  ,\quad  \matr T  = \matr T^\top, \label{eqn:divT}
\end{equation}
with $\vec b$ a density of applied body loads (per unit current volume), typically self-weight in our context. Henceforth, we focus on the case $\vec b=\vec 0$.  
The symmetry of the stress tensor results from the absence of applied torque and micropolar structure in the material. In a more general theory, asymmetric stresses may be introduced to reflect the possible local twist of the cells \citep{CHAKRABORTY2021110736,wada2012hierarchical}.

\subsubsection{Balance of internal energy}


To complete the theory, we discuss the thermodynamics of the system. The balance of energy for the total internal energy $\mathcal E$ of a region $\Omega \subset \mathcal B$ encompasses internal work contributions $\mathcal P$, heat $\mathcal Q$, and addition of new material $\mathcal S$:
\begin{equation}
    \fdiff{\mathcal E }{t}  = \mathcal P + \mathcal Q + \mathcal S  ,\label{eqn:global-energy}
\end{equation}
where
\begin{equation}
    \mathcal E = \int_\Omega e \,\diff v,
\end{equation}
with $e$ the internal energy per unit volume of mixture.
%
The work rate encompasses the work of internal forces,  the works required to make room for new fluid, osmolyte and  solid material ($w_s$) during growth, and the work $\bar w_o$ accounting for the active transport of the osmolytes, reflecting chemical processes that we do not address explicitly. Altogether, we have
\begin{align}
    \mathcal P =&  \int_{ \Omega}  { \vec T  : \matr L}\, \diff {v}   + \int_\Omega p  \xi_f \, \diff v- \int_{\partial\Omega} p  \vec j_f \cdot\diff {\vec a}+ \int_\Omega p  \xi_o \, \diff v- \int_{\partial\Omega} p \vec j_o \cdot\diff {\vec a}  + \int_\Omega    w_s \xi_s \,  \diff v + \int_{  \Omega }  \bar w_o \diff {v}   , 
\end{align}
where $\vec j_f = \phi_f \vec w_f$ and $\vec j_o = \phi_o \vec w_o$ are the relative fluxes of fluid and osmolytes through the mixture; with $\vec w_f \eqdef \vec v_f - \vec v$ the seepage velocity of the fluid and $\vec w_o \eqdef \vec v_o - \vec v$; and where $\diff {\vec a}$ and $\diff v$ denote respectively the outward normal surface element and the volume element for the integration.
The heat contribution encompasses a bulk source $r$ and a flux $\vec q$ through the boundary:
\begin{equation}
    \mathcal Q = \int_\Omega r \diff v - \int_{\partial \Omega} \vec q\cdot \diff{\vec a}.
\end{equation}
Energy is also added to the system through addition of new  material. We note $e_f$, $e_o$ and $e_s$ the internal energy of the fluid, the osmolytes and the solid so that
\begin{equation}
    \mathcal S = \int_\Omega e_s \xi_s \, \diff v + \int_\Omega e_f \xi_f \diff{v} - \int_{\partial \Omega} e_f \vec j_f \cdot \diff{\vec a}+\int_\Omega e_o \xi_o \diff{v}- \int_{\partial \Omega} e_o \vec j_o \cdot \diff {\vec a}    .
\end{equation}
Using the divergence theorem to eliminate the boundary terms, then the Leibniz rule for differentiation under the integral sign, and the localisation theorem,
we derive the local form of the balance of internal energy 
\begin{equation}\label{eqn:conservation-energy}
     { \pdiff{e}{t}   } + \divergence  ( e \vec v  ) = \vec T  : \matr L + r   + h_f \xi_f  +  h_o \xi_o +  h_s \xi_s+ \bar w_o- \divergence\pp{  h_f  \vec j_f + h_o \vec j_o + \vec q },
\end{equation}
where we have introduced $h_f =  e_f + p $, $h_o =  e_o + p $ and $h_s = e_s + w_s $ the enthalpies per unit volume of pure fluid, osmolytes and solid material, respectively. Alternatively, it is convenient to express the balance of energy in reference configuration as
\begin{equation}\label{eqn:conservation-energy-refconfig}
    \dot E^{0} = \vec P  : \dot {\matr F} + R^{0}  +  {h_f \Xi_f^0 +   h_o \Xi_o^0 +  h_s \Xi_s^0   + \bar W_o^0}- \Div \pp{  h_f \vec J_f^0 + h_o \vec J_o^{0} + \vec Q^{0} },
\end{equation}
where $\Div$ denotes the Lagrangian divergence;  
$E^0 = Je$ is the internal energy per unit reference volume; 
$\matr P = J\matr T\matr F^{-\top}$ is the Piola–Kirchhoff stress \citep{holzapfel2000nonlinear}; and $\vec Q^{0} = J\matr F^{-1} \vec q$, $\vec J_o^{0} = J\matr F^{-1} \vec j_o$ and $ \vec J_f^0 = J \matr F^{-1} \vec j_f$ are the flux terms pulled back to the reference configuration. We define the quantities $\Phi_f^0 = J\phi_f$, $\Phi_o^0 = J\phi_o$,  and $\Phi_s^0 = J\phi_s=J_G\Phi_s$, measuring respectively the  {current} volume of each component per unit {reference} volume, so that $\Xi_s^0=\Phi_s^{0} \Gamma$, $\Xi_o^0 = J\xi_o$ and $\Xi_f^0 = J\xi_f$ are the local production of solid, osmolytes and fluid  volume per unit reference volume; $R^0 = Jr$ is the heat source per unit reference volume; $\bar W_o^0 = J\bar w_o$ is the work due to active transport per unit reference volume.

\subsubsection{Imbalance of entropy}

The second law of thermodynamics is expressed through the Clausius-Duhem inequality, expressed in terms of the entropy density $S^0$ (entropy per unit initial volume) as \citep{XUE2016409,Goriely2017}
\begin{align} \label{eqn:clausius-duhem}
      \dot S^0 &\geq     \frac{R^0  }{\theta}+ s_f \Xi_f^0 + s_o\Xi_o^0    + s_f\Xi_s^0  + \bar S_o^0 - \Div\pp{ \frac{\vec Q^0  }{\theta} +  s_f   \vec J_f^0 + s_o \vec J_o^0 } ,  
\end{align}
with $\theta$ denoting the thermodynamic temperature; and where $s_f$, $s_o$ and $s_s$ are the volumetric entropy densities of the added fluid, osmolyte and solid materials, respectively; and $\bar S_o^0$ denotes the entropy contribution from active transport. Henceforth, we assume isothermal conditions for simplicity, so we take $\vec Q^0 = \vec 0$ and $R^0=  0$.  Introducing the Helmholtz free energy
$
    \Psi^0 = E^0 -  \theta S^0
$ and applying the energy balance equation \eqref{eqn:conservation-energy-refconfig}, we can reformulate the previous inequality as
\begin{align}
  \dot\Psi^0  \leq    \vec P  : \dot{\matr F} + g_f \Xi_f^0  +  g_o \Xi_o^0 +  g_s \Xi_s^0  + \bar G_o^0 - \Div\pp{  g_f \vec J_f^0 + g_o \vec J_o^0 }  ,\label{eqn:ineq-Psi0}
\end{align}
where $g_s = h_s - \theta  s_s$, $g_f = h_f - \theta s_f$ and $g_o = h_o - \theta s_o $ denote the Gibbs free energy densities of the solid, fluid and solute material, respectively; and $\bar G_o^0 \eqdef \bar W_o^0 - \theta  \bar S_o^0$. As detailed in the next section, the inequality \eqref{eqn:ineq-Psi0} is useful to derive constitutive laws for growth and material transport. 

\subsection{Constitutive laws\label{constitutive}}


\subsubsection{Coleman-Noll procedure}
To close the system we need to formulate constitutive laws for the material. Therefore, we follow the approach of \cite{XUE2016409} and apply the Coleman-Noll procedure to the dissipation inequality \eqref{eqn:ineq-Psi0} to constrain thermodynamically the constitutive laws  \citep{Coleman1963167,gurtin2010mechanics,Goriely2017}. Firstly, we decompose the free energy $\Psi^0$ into a solid part $\Psi_s^0 $ and a fluid part $\Psi_f ^0 =  \Phi_f^0 g_f +\Phi_o^0  g_o   - \Phi^0  p $ \citep{coussy2004poromechanics}; so that $\Psi^0 = \Psi_s^0 +\Phi_f^0 g_f +\Phi_o^0  g_o   - \Phi^0  p$.  For an inherently incompressible solid component, the porosity $\Phi ^0 = J_G (J_A - \Phi_s )$ is determined by the deformation. 
We introduce the mechanical free energy of the material in the form 
%
\begin{equation}\label{eqn:W0def}
    W^0 (\matr A, \matr G, p) 
    = \Psi_s^0(\matr A, \matr G)  - p J_G  (J_A - {\Phi_s}),
\end{equation}
encompassing an elastic contribution and a hydrostatic contribution. 
The energy per unit intermediate volume is
\begin{equation}\label{eqn:Wdef}
    W(\matr A, p) = J_G^{-1} W^0 (\matr A, \matr G, p) = \Psi_s (\matr A)   - p    ( J_A - \Phi_s), 
\end{equation}
where $\Psi_s(\matr A) = J_G^{-1}\Psi_s^0 (\matr A, \matr G) $ defines the strain energy density of the solid, per unit intermediate volume.
%
Then, combining \eqref{eqn:ineq-Psi0,eqn:W0def} with the balance of mass relations $\dot\Phi^0_f  =    \Xi_f^0 - \Div \vec J_f^0 $ and $\dot\Phi^0_o  = \Xi_o^0 - \Div \vec J_o^0 $ \eqref{eqn:balance-osmolyte-mass,eqn:massf}, and the Gibbs-Duhem equality $\dot p\Phi^0 =  \dot g_f\Phi_f^0 +\dot g_o \Phi_o^0 $, 
we obtain 
\begin{align}\label{eqn:clausisu-duhem-W0}
  \dot W^0  \leq    \vec P  : \dot{\matr F}    - \dot{p} \Phi_f^0 +  
  \bar G_o^0 - \vec J_f^0 \cdot \Grad g_f - \vec J_o^0 \cdot \Grad g_o   .
\end{align}
Finally, expanding $\dot W^0$ as
$
    \dot W_0 J_G ^{-1} =  {W\matr G^{-\top}:\dot {\matr G} + \pp{\linepdiff{W}{\matr A}} : \dot{\matr A} + \pp{\linepdiff{W}{p} }\dot p}
$
and substituting into \eqref{eqn:clausisu-duhem-W0} using \eqref{eqn:W0def,eqn:Wdef} provides
\begin{align}\label{eqn:coleman-noll}
0  \leq &   \pp{\matr P \matr G^\top  -  J_G\pdiff{W}{\matr A}}:\dot{\matr A}     +   \pp{\matr A^\top\matr P  + \pp{g_s \Phi_s^0 - J_G W} \matr G^{-\top} }:\dot{\matr G}   
+ \bar G_o^0 - \vec J_f^0 \cdot \Grad g_f - \vec J_o^0 \cdot \Grad g_o   .
\end{align}

By the argument of \cite{Coleman1963167}, the inequality \eqref{eqn:coleman-noll} must hold for any admissible process. In particular, since $\dot{\matr A}$ is arbitrarily prescribed, the following standard equalities must hold universally
    \begin{equation}\label{eqn:coleman-noll-stress}
    \matr P  =  \pdiff{W^0}{\matr A} \matr G^{-\top} \quad \Leftrightarrow \quad \matr T = J_A^{-1}\matr A \pdiff{W}{\matr A} 
\end{equation}
which provides the standard stress-strain relation for a hyperelastic material. 
 The resulting dissipation inequality,
\begin{align}\label{eqn:coleman-noll2}
0  &\leq       \pp{\matr A^\top\matr P  + \pp{g_s \Phi_s^0 - J_G W} \matr G^{-\top} }:\dot{\matr G} 
+ \bar G_o^0 - \vec J_f^0 \cdot \Grad g_f - \vec J_o^0 \cdot \Grad g_o ,
\end{align}
highlights the different modes of dissipation in the system, coming from distinct biophysical processes, specifically growth and transport. Thus, these contributions must  satisfy \eqref{eqn:coleman-noll2} individually, hence
\begin{subequations}
\begin{equation}\label{eqn:coleman-noll-growth}
   \pp{  g_s \Phi_s \matr 1- \matr B}: \matr L_G  \geq 0 ,
\end{equation}
\begin{equation}\label{eqn:coleman-noll-flux}
   \bar   G_o^0 -\vec J_f^0 \cdot \Grad g_f    - \vec J_o^0 \cdot \Grad g_o  \geq 0.
\end{equation}
\end{subequations}
Here, $\vec B  \eqdef  W\matr 1-J_A \matr A^\top\matr T\matr A^{-\top} $ denotes the Eshelby stress tensor, which appears as an important quantity for growth \citep{epstein2000thermomechanics,vandiver2009morpho}.


\subsubsection{Growth law\label{growth-law}}

We discuss the constitutive law for the growth of the cell matrix.  Focusing on the solid component, we first decompose the stress additively into a solid and fluid part as $ \matr T =  \matr T_s   -  \phi p  \matr 1$ where $\matr T_s$ is the partial stress due to the solid component \citep{coussy2004poromechanics,preziosi2002darcy}:
\begin{equation}\label{eqn:partial-stress}
     \matr T_s 
  =  J_A^{-1}\matr A \pdiff{\Psi_s}{\matr A} - \phi_s p \matr 1.
\end{equation}
We can  rewrite \eqref{eqn:coleman-noll-growth2} in terms of the partial Eshelby stress $\matr B_s  \eqdef  \Psi_s\matr 1 -   \phi_s J_A \matr A^\top\matr T_s\matr A^{-\top}
$ ($=\matr B$) as
\begin{equation}\label{eqn:coleman-noll-growth2}
    \pp{g_s\Phi_s\matr 1 -  \matr B_s}:\matr L_G \geq 0 .
\end{equation}
This dissipation inequality expresses a thermodynamic constraint on the growth law. In particular, the case $g_s=0$, where the solid is added with no energy to the matrix, defines a \textit{passive} growth process \citep{Goriely2017}. 
Generally, from \eqref{eqn:coleman-noll-growth2}, a natural path is then to adopt the Eshelby stress $\matr B_s$ as a driving force for growth. A common approach then is to postulate the existence of a homeostatic stress $\matr B^*$ such that $\matr B^*:\matr L_G = \Phi_sg_s\Gamma$, so a natural growth law can be chosen of the general form $\matr L_G = \mathcal H :  (\matr B^* - \matr B_s)$ with  $\mathcal H$ a fourth-order extensibility tensor such that the inequality holds for any $\matr B_s$ \citep{ambrosi2007stress,DICARLO2002449,XUE2016409,dunlop2010theoretical,tiero2016morphoelastic,ambrosi2012interplay}. In plants however, it is unlikely that such homeostatic stress is actively maintained. Instead, the growth of  plant cell walls is generally seen as a passive, plastic-like process \citep{ali_force-driven_2016,dyson_model_2012,barbacci2013another} which involves complex  anisotropies and nonlinear threshold effects  \citep{oliveri2019regulation,boudon_computational_2015}. Whether simple growth laws based on the Eshelby stress can reproduce  basic properties of plant matter remains unclear to us. 
Thus, we reserve this problem for another occasion and instead adopt a simpler, phenomenological approach. Indeed, \textit{strain}-based growth laws  of the form $\matr L_G = \mathcal G(\matr A)$ have been adopted \citep{boudon_computational_2015,bozorg_continuous_2016,mosca2024growth,zhao2020microtubule,KellyBellow2023,rojas_chemically_2011,laplaud2024assessing}, where typically, the cell walls expand once they surpass a certain strain threshold. These growth laws are relatively simple to parameterise and capture elegantly some observed phenomenological properties of plant matter, for example the commonly-accepted fact that growth is slower in rigid tissues or along the direction of load-bearing cellulose microfibrils (all other things being equal). 

Following previous strain-based models we postulate a strain-based growth law where we assume that during growth, the strain is maintained close to a strain threshold $\matr E^*$. Therefore, we posit a linear growth law of the form
\begin{equation}\label{eqn:strain-based-growth}
    \mathcal G(\matr A) =   \mathcal K:\ramp{\matr E - \matr E^*},
\end{equation}
with $\mathcal K$ a fourth-order tensor controlling the rate and anisotropy of the growth (with unit inverse time); $\matr E \eqdef (\matr A^\top\matr A - \matr 1)/2$ the symmetric Lagrangian elastic tensor; $\matr E^*$ is a threshold strain tensor; and $\ramp{\cdot}$ is an invariant extension of the ramp function to symmetric tensors, that is, for any symmetric tensor $\matr S$ with eigenvalues $S_i$ and corresponding normalised eigenvectors $\vec s_i$, we define
$
    \ramp{\matr S} \eqdef \max\pp{0,S_i} \,\vec s_i \otimes \vec s_i$.
For comparison, \eqref{eqn:strain-based-growth} is a generalisation of the growth law introduced by \cite{boudon_computational_2015}. 

Note that  
there is no guarantee anymore that the dissipation inequality \eqref{eqn:coleman-noll-growth2},
\begin{equation}\label{eqn:passive-growth}
      g_s \Phi_s  \geq  \frac{\matr B_s :\mathcal G(\matr A)}{\Tr\pp{\mathcal G(\matr A)}},
\end{equation}
will be always satisfied with $g_s = 0$. 
(However, in practice, the r.h.s. appears to be indeed negative in many cases studied here.) 
We conclude that strain-based growth processes may not be universally passive. 
We note that, recently, some authors have proposed that material insertion and pectin expansion may also drive growth, independently of turgor \citep{haas2020pectin,HAAS2021100054}. This hypothesis, albeit contentious \citep{cosgrove2020plant}, implies that growth in plant cell walls may be--at least partly--an active process.

\subsubsection{Transport\label{transport}}
Next, we discuss the second inequality \eqref{eqn:coleman-noll-flux} for the material fluxes. Using $\vec J_o^0 = J\phi_o  \matr F^{-1} \vec w_o$ and $\vec J_f^0 = J\phi_f \matr F^{-1} \vec w_f$ and postulating an active transport contribution of the form $\bar g_o = \phi_o \bar{\vec f}_o \cdot \vec w_o $ with $\bar {\vec f}_o$ an effective force sustaining active transport, \eqref{eqn:coleman-noll-flux} becomes 
\begin{equation}\label{eqn:coleman-noll-flux2}
   -\phi_f\vec w_f  \cdot \grad g_f  + \phi_f \vec w_o  \cdot ( \bar{\vec f}_o - \grad g_o ) \geq 0.
\end{equation}
To make progress, we introduce Onsager's reciprocal relations \citep{onsager1931reciprocal,martyushev2006maximum,XUE2016409}
\begin{subequations}\label{eqn:onsager}
\begin{equation}
    - \phi_f \grad g_f = L_{fo} (\vec v_f - \vec v_o) + \matr L_{fs} (\vec v_f - \vec v) = - L_{fo}   \vec w_o + ( \matr L_{fs} + L_{fo}\matr 1) \vec w_f,
\end{equation}
\begin{equation}
      \phi_o\bar{\vec f}_o - \phi_o \grad g_o = L_{of} (\vec v_o - \vec v_f) + \matr L_{os} (\vec v_o - \vec v) = - L_{of}  \vec w_f + ( \matr L_{os} + L_{of}\matr 1) \vec w_o , 
\end{equation}
\end{subequations}
where $L_{of}=L_{fo}$ is a fluid-osmolyte drag coefficient; and $\matr L_{os}$ and $\matr L_{fs}$ are positive-definite symmetric second-order tensors that characterise the anisotropic osmolyte-solid and fluid-solid drags, respectively, to reflect the anisotropic structure of the solid matrix. Indeed, substituting \eqref{eqn:onsager} into \eqref{eqn:coleman-noll-flux2} gives
$
   \vec w_f \cdot (\matr L_{fs} \vec w_f ) +    \vec w_o \cdot (\matr L_{os} \vec w_o ) +   L_{fo}(\vec w_f - \vec w_o)^2 \geq 0$, 
which is always satisfied. These relations may be viewed as a generalisation of Fick's law to several transported species.
Inverting \eqref{eqn:onsager} and using the Gibbs-Duhem equality $\phi \grad p = \phi_f \grad g_f + \phi_o \grad \phi_o$  we obtain the relative velocities
\begin{subequations}
\begin{equation}
      \vec w_f= \bb{\matr L_{os}\matr L_{fs} + L_{of} (\matr L_{os}+\matr L_{fs}) }^{-1}\bb{\phi_o\pp{L_{fo}    \bar{\vec f}_o +\matr L_{os}   \grad \phi_o } -\phi \pp{\matr L_{os} + L_{of}\matr 1}  \grad p} ,
\end{equation}
\begin{equation}
    \vec w_o = \pp{\matr L_{os} + L_{of}\matr 1}^{-1}\pp{\phi_o\bar{\vec f}_o - \phi_o \grad g_o +  L_{fo} \vec w_f}.
\end{equation}
\end{subequations}
In particular, in the limit of high osmolyte-solid drag $\matr L_{os} \gg \matr L_{fs},  L_{of}$, we derive
\begin{equation}\label{eqn:high-drag-w}
    \vec w_f \approx  \pp{\matr L_{fs} + L_{of} \matr 1 }^{-1} \pp{\phi_o     \grad \phi_o   - \phi    \grad p}, \quad 
   \vec w_o  \approx \phi_o\matr L_{os}  ^{-1}\pp{\bar{\vec f}_o -  \grad g_o}.
\end{equation}
Biologically, this assumption expresses the idea that in the absence of active transport, the osmolytes remain mostly confined within the cells (with some slow diffusive leakage) and are not convected by the fluid. Following \cite{XUE2016409}, the drag coefficients are given as
$
    \matr L_{fs} = \phi \matr K^{-1}$ and $ L_{of} =  {R_g\theta c_o}/{D_o}
$,
with $\matr K $ the second-order symmetric \textit{permeability tensor} that characterises the permeability of the mixture to the fluid (expressed in unit area per pressure per time); $c_o$ is the molar concentration of osmolytes; $D_o$ is the diffusivity of the osmolytes in the fluid; and $R_g$ the universal gas constant.  
The  chemical potential $g_o$ of the solutes can be related to their molar concentration $c_o$ in the fluid through
\begin{equation}\label{eqn:chemical-potential}
    g_o = g_o^0 + \frac{R_g\theta}{v_o}\ln \pp{ {c_o}/{c_o^0}}  ,
\end{equation}
with $v_o$ the molar volume of the osmolytes; and $g_o^0$ and $c_o^0$ are reference values, taken to be constant.
Using the formulae $\vec j_f=\phi_f\vec w_f$ and $c_o = \phi_o / v_o\phi $ with \eqref{eqn:chemical-potential,eqn:high-drag-w}, we obtain finally

%
\begin{equation}
    \vec j_f \approx \phi_f \pp{ \matr K^{-1} + \frac{R_g\theta c_o}{D_o\phi} \matr 1 }^{-1}  \grad\pp{R_g\theta c_o   -  p}.
\end{equation}
For small concentration, more specifically for $R_g\theta c_o  \ll  D_o\phi \matr K^{-1}$, we recover a Darcy-type law
\begin{equation}
    \vec j_f \approx  - \phi_f\matr K \grad \psi, \label{eqn:darcy}
\end{equation}
where $\psi = p - \pi$ as in \eqref{eqn:ortega-fluxes}; with the osmotic pressure $\pi$ given by the van 't Hoff relation $\pi \eqdef R_g\theta c_o$.  Combining \eqref{eqn:phisdeformation,eqn:sum-balance-mass,eqn:darcy,eqn:saturation} we can finally derive
\begin{equation}\label{eqn:masss-final}
    \divergence\bb{  \vec v   + \vec j_o -\pp{1 - \phi_o -  {\Phi_s}/{J_A}} \matr K \grad \psi  }  = \xi_s  + \xi_o + \xi_f.
\end{equation}

In our context, a useful simplification comes from the fact that, in plants, the fluid content actually accounts for most of the volume of the mixture, thus, water mass balance is barely affected by the small volume of the osmolyte and the cell walls. Therefore, we next take $ \xi_s \ll \xi_f  $, $\vec j_o \ll \vec j_f$, $\phi_o, \phi_s\ll \phi$ and $\matr T_s \approx \matr T + p\matr 1$.

\subsection{A closed system of equations}
Combining \eqref{eqn:balance-osmolyte-mass,eqn:masss-final,eqn:strain-based-growth,eqn:multiplicative,eqn:divT}, we obtain the following system of equations:
\begin{subequations}\label{eqn:complete-set-equations}
\begin{equation}
    \divergence  \pp{\vec v-\matr K \grad \psi } =    \xi_f ,\label{eqn:masss-final2}
\end{equation}
\begin{equation}\label{eqn:masso-final2}
     \pdiff{\phi_o}{t} + \divergence\pp{\phi_o\vec v + \vec j_o}  =  \xi_o,
\end{equation}
\begin{equation}
    \divergence \matr T_s  + \vec b= \grad p,
    \label{eqn:balance-linear-momentum2}
\end{equation}
\begin{equation}
      \matr T_s  = \matr T_s ^\top ,\label{eqn:balance-angular-momentum2}
\end{equation}
\begin{equation}
    \matr F = \matr A \matr G ,\label{eqn:multiplicative2}
\end{equation}
\begin{equation}
    \matr L_G  =  \mathcal K:\ramp{\matr E - \matr E^*},
\end{equation}
\begin{equation}\label{eqn:terzaghi2}
    \matr T_s  = J_A^{-1} \matr A \pdiff{\Psi_s}{\matr A}, 
\end{equation}
\begin{equation}\label{eqn:osmotic-potential}
    \psi = p -   R_g\theta \phi_o / v_o.
\end{equation}
\end{subequations}
In total, we have a closed system of thirty-three partial-differential equations for thirty-three variables: the nine components of $\matr A$, the nine components of $\matr G$, the three components of $ \boldsymbol \chi$, the nine components of $\vec T_s$, and the three scalar variables $p$,  $\psi$ and $\phi_o$.
\subsection{Boundary conditions}

This system must be equipped with appropriate boundary conditions. Defining the outer normal $\vec n$  to the boundary $\partial \mathcal B$, typical boundary conditions encountered in our context describing the flux of water through the boundary are Robin boundary conditions of the form
\begin{equation}
 \vec j_f \cdot \vec n = -\pp{\matr  K \grad \psi }\cdot \vec n=  k (\psi^*  - \psi)   \quad \text{on $\partial \mathcal B$},\label{eqn:robin}
\end{equation}
expressing a flux across the boundary due to a difference of water potential with the outside $\psi^*$, where $k$ is the interfacial hydraulic conductivity. No-flux boundary conditions $
    \pp{\matr  K \grad \psi }\cdot \vec n =  0 $
 are a particular case of \eqref{eqn:robin}. 
Note that when $k\rightarrow\infty$, the condition \eqref{eqn:robin} simplifies to a Dirichlet constraint
\begin{equation}\label{eqn:dirichlet}
  \psi  = \psi^* \quad \text{on $\partial \mathcal B$}.
\end{equation}
Similarly, the mass balance equation for the osmolytes may be equipped with Dirichlet (fixing $c_o$ at the boundary) or flux boundary condition. The usual boundary condition for the stress \eqref{eqn:balance-linear-momentum2} is
\begin{equation}\label{eqn:bc-stress}
    \matr T\vec n = \vec t_0  \quad \text{on $\partial \mathcal B$} ,
\end{equation}
with $\vec t_0$ the applied traction density.

\subsection{Apparent elasticity of a growing tissue\label{apparent-elasticity}}

Given a solution to the system, an adscititious problem is to characterise the effective elasticity of the turgid tissue subject to residual stresses. Indeed, at timescales shorter than that of growth and water transport, water is effectively trapped in the tissue and the mixture can then be seen as a macroscopically \textit{incompressible} hyperlastic material. Given a field of  elastic pre-deformation $\matr A$ and pressure $p$, the effective strain energy function of the mixture with respect to an incremental, superimposed deformation $\hat{\matr A}$ is 
\begin{equation}\label{eqn:effective-energy}
    \hat W( \hat {\matr A}, \hat p) = J_A^{-1} W ( \hat{\matr A}{\matr A}, p )   - \hat  p (\det\hat{\matr A} - 1 ) , 
\end{equation} where $ \hat   p$ is an undetermined Lagrange multiplier that accounts for the incompressibility constraint $\det\hat{\matr A} = 1$. The associated Cauchy stress is then
\begin{equation}\label{eqn:effective-stress}
    \hat{\matr T} = \hat{\matr A} \pdiff{\hat W}{\hat {\matr A}} 
    = \matr T (\hat{\matr A} {\matr A},p) -   \hat p  \matr 1.
\end{equation}
 This expression is useful to describe the overall elastic response of the system to external forces, e.g. in the context of compression experiments performed on entire tissues \citep{hamant2008developmental,pieczywek2017compression,robinson2018global,zhu2003mechanics}, or to explore mechanical stability under growth-induced differential stresses or external loads \citep{BenAmar2005,vandiver2008tissue}. 

\section{Longitudinal growth\label{rod}: hydraulic competition and growth-induced water gradients}


\subsection{General problem}

To illustrate the behaviour of the system, we first examine the simple scenario of a growing, straight cylindrical rod so we reduce the system \eqref{eqn:complete-set-equations} to one dimension. We define $s$, $S $ and $S_0 $, the arc lengths in the current, intermediate and initial configurations respectively. The associated total, growth and elastic stretches are $\lambda \eqdef \linepdiff{s}{S_0}$, $\gamma  \eqdef \linepdiff{S}{S_0}$ and $\alpha\eqdef\linepdiff{s}{S}$, with $
    \lambda=\alpha\gamma
$ \eqref{eqn:multiplicative}. From \eqref{eqn:rates,eqn:velocity-gradient}, we derive
\begin{equation}
    \pdiff{v}{s} = \frac{\dot{\alpha}}{\alpha} + \frac{\dot{\gamma}}{\gamma},
\end{equation}
with $v$ the longitudinal Eulerian velocity (positive towards increasing $s$).
Assuming that the rod is unloaded, the  balance of linear momentum \eqref{eqn:balance-linear-momentum2} for the axial partial stress $t_s$ yields $t_s=p$. 
System \eqref{eqn:complete-set-equations} then reduces to
\begin{subequations}
\label{eqn:1d-eulerian}
\begin{equation}\label{eqn:balance-1d-mass}
    \pdiff{}{s}\pp{v  - {K \pdiff{\psi}{s}}} =  
    \xi_f , 
\end{equation}
\begin{equation}\label{eqn:balance-1d-osmolyte-mass}
     \pdiff{\phi_o}{t} + \pdiff{}{s}\pp{\phi_o  v +   j_o}  =  \xi_o,
\end{equation}
\begin{equation}
    \pdiff{v}{s} =    \frac{1}{2 \tau}\ramp{\alpha^2 - {\alpha^*}^2} +  \pdiff{\alpha}{t} + \frac{v}{\alpha}\pdiff{\alpha}{s} ,\label{eqn:continuum-ortega}
\end{equation}
\begin{equation}
    p = \pdiff{\Psi_s}{\alpha},
\end{equation}
\begin{equation}\label{eqn:1d-osmotic-potential}
    \psi = p -  {R_g\theta \phi_o }/{v_o},
\end{equation}
\end{subequations}
where $\tau$ is the characteristic time of wall synthesis during growth (i.e. $\mathcal K = 1/\tau$); $\matr K = K$ is the permeability coefficient; and $\vec j_0=j_0$ is the osmolyte flux. 

For simplicity, we focus in this section on infinitesimal elastic deformations and define $\epsilon = \alpha - 1$ the \textit{infinitesimal strain}, with $\epsilon \ll 1$, and the strain threshold $\epsilon^* = \alpha^* - 1\geq 0$. Thus, Hooke's law provides
\begin{equation}
    \label{eqn:pepsilon}  p  = E \epsilon, 
\end{equation}
with $E$ the effective Young's modulus of the solid. 
Here we  assume that the osmotic pressure $\pi$ is maintained constant w.r.t $s$ and $t$, so we elide  \eqref{eqn:1d-osmotic-potential,eqn:balance-1d-osmolyte-mass} and substitute $\linepdiff{\psi}{s}=\linepdiff{p}{s}$ into \eqref{eqn:balance-1d-mass}. Similarly, we take $K$,  $\epsilon^*$, and $\tau$ constant and homogeneous. Indeed, the focus here is on the role of a spatially heterogeneous stiffness, which has been linked experimentally \citep{kierzkowski_elastic_2012} and theoretically \citep{cheddadi2019coupling, alonso2024water,boudon_computational_2015} to organ patterning in development. Therefore, we allow the rigidity to vary spatially as $E(s,t) = E_0 f(s)$ with $f$ a dimensionless function of order unit, and $E_0$ a characteristic Young's modulus of the rod. Finally, taking $\tau$, $E_0$ and $\sqrt{KE_0\tau}$ as reference time, pressure and length, respectively, we can nondimensionalise the system through the substitutions
\begin{equation}
    \tilde v =v \sqrt{\frac{\tau}{KE_0}},  \quad \tilde s = \frac{s}{\sqrt{KE_0\tau}},\quad \tilde   t = \frac{t}{\tau}, \quad \tilde \xi_f =  \xi_f \tau, \quad \tilde p =  \frac{p}{E_0}.
\end{equation}
In particular, the length $\mathcal L_h \eqdef\sqrt{KE_0\tau}$ defines the characteristic  \textit{hydromechanical length} of the system.   Altogether, we obtain to $O(\epsilon)$:
\begin{subequations}\label{eqn:1d-eulerian-nondim}
\begin{equation}
  \pdiff{\tilde v}{\tilde s}  =   \ppdiff{\tilde p}{\tilde s}  
  + \tilde \xi_f, \label{eqn:balance-mass-linear1d}
\end{equation}
\begin{equation}
    \pdiff{\tilde v}{\tilde s} =   \ramp{\epsilon- \epsilon^*} +  \pdiff{\epsilon}{\tilde t} +\tilde v \pdiff{\epsilon}{\tilde s} ,
\end{equation}
\end{subequations}
with $\tilde p = f  \epsilon$. 
To simplify notations, we drop the tildes and use nondimensionalised variables. For comparison, this system ressembles the static model of \cite{plant1982continuum}, and extends the Lockhart-Cosgrove-Ortega model \eqref{eqn:ortega,eqn:ortega-fluxes} to a one-dimensional continuum. Note also that the r.h.s of \eqref{eqn:passive-growth} is always negative or zero here, thus, by the argument of \cref{growth-law}, our strain-based growth law is compatible with a passive growth process. In the next two sections, we study the effect of heterogeneous elastic moduli on the growth behaviour of the rod.

\subsection{Material heterogeneity and hydraulic competition\label{competition}}

Here we are interested in the boundary effects arising between two regions with different mechanical properties, specifically the effect of different stiffnesses. Therefore, we consider a rod with initial arclength $S_0\in \bb{-L_0/2,L_0/2}$, divided in the middle into two part of different rigidity, that is, we posit the rigidity field given by 
\begin{equation}
   f (s(S_0)) =   1 - \eta  \theta(S_0),
\end{equation}
where $  0 \leq \eta < 1$ denotes the  stiffness step at the interface $s=0$; and with
$\theta$ the Heaviside step function (i.e. $\theta(x) = 0$ if $x<0$ and  $\theta(x) = 1$ if $x\geq 0$). The region $S_0>0 $ is softened, with stiffness $1-\eta $, with respect to the base stiffness equal to unity. Assuming $s(0)=v(0)=0$, without loss of generality (i.e. considering an observer located at the interface), we have $f(s)= 1 - \eta  \theta(s)$. Here we assume no-flux boundary conditions at both ends of the domain, however we allow water entry through a bulk source $\xi_f = k (p^* - p)$, with $k$ the effective permeability with the outside, and $p^*$ a constant base effective pressure, encompassing the excess osmolarity relative to the outside, and the outer hydrostatic pressure.

To gain insight into the shape of the solutions, we focus on the steady regime, i.e. on self-similar stationary solutions  on an infinite line. On setting $\linepdiff{\epsilon}{t}=0$ in \eqref{eqn:1d-eulerian-nondim}, the problem reduces to:
\begin{subequations}\label{eqn:self-similar}
\begin{equation}
v'  =   p''  
  + k(p^* - p), \label{eqn:self-similar1}
\end{equation}
\begin{equation}
   v' =   \ramp{\epsilon- \epsilon^*}    +  v  \epsilon', \label{eqn:self-similar2}
\end{equation}
\end{subequations}
with the apostrophe denoting derivative with respect to $s$, and where $p=f\epsilon$. 
To make progress, we consider a small perturbative softening  $\eta\ll 1$. Therefore we expand all variables as power series of $\eta$, i.e. $p = p_0 + \eta p_1 + \dots$, $\epsilon = \epsilon_0 + \eta \epsilon_1  + \dots$ and $v = v_0 + \eta v_1  + \dots$,  
and then treat each order in \eqref{eqn:self-similar} separately. To ease calculations, it is also convenient here to ignore the threshold effect in \eqref{eqn:self-similar2} and assume $p^*\geq\epsilon\geq\epsilon^*$. The base solution at $O (1)$ has uniform pressure and velocity gradient and is given by
\begin{subequations}
\begin{equation}
   p_0= \epsilon_0 = \frac{\epsilon^* + kp^*}{1+k},
\end{equation}
\begin{equation}
    v_0 = a s, \quad \text{with} \quad a = \frac{ p^*-\epsilon^* }{1+k^{-1}} .\label{eqn:base-velocity}
\end{equation}
\end{subequations}
Unsurprisingly, $p_0$ can be identified to the Lockhart pressure \eqref{eqn:lockhart-pressure}. At $O\pp{\eta }$, we have
\begin{equation}
 v_1'  =   p_1'' - k   p_1, 
\quad
   v_1'= \epsilon_1  +  v_0   \epsilon_1' .
\end{equation}
Eliminating $v_1'$ and using $  p_1 =     \epsilon_1 - \epsilon_0 \theta(s)$ and $p_0=\epsilon_0$, we obtain a single equation for $p_1$,
\begin{equation}
 p_1'' - as    p_1' -\pp{1+k}  p_1   = p_0\theta(s), \label{eqn:p1} 
\end{equation}
defined on both regions $s<0$ and $s>0$. A general solution to this equation can be expressed in terms of the Kummer confluent hypergeometric function $\, _1F_1$, the gamma function $\Gamma$ and the Hermite polynomials (of the first kind) $H_\lambda$. By imposing the condition that pressure must be bounded and continuously differentiable at $s=0$, we can determine the four integration constants and obtain the compound asymptotic solution  
    \begin{equation}
   p  (s) = p_0 + \frac{\eta p_0 }{k+1}   \bb{\frac{\pi  \csc \left( \pi c\right)}{\Gamma \left( c/{2 }\right) \Gamma \left(1-c\right)} H_{-c}\left(s\sqrt{ {a}/{2}}\right)- \theta(s) - \theta(-s) h_{-c}\pp{s\sqrt{ {a}/{2}}}  }+ O (\eta ^2),\label{eqn:asymptotic-pressure}
    \end{equation}
    with $c \eqdef \pp{k+1}/a  $; and $h_\lambda = \, _1F_1\left(- \lambda/2; {1}/{2};x^2\right)$  the Hermite functions ($\pi$ denotes here the ratio of a circle's circumference to its diameter). The expansion rate $v'$ is obtained directly by substituting this expression into \eqref{eqn:self-similar1}.

\begin{figure}[ht!]
    \centering
\includegraphics[width=\linewidth]{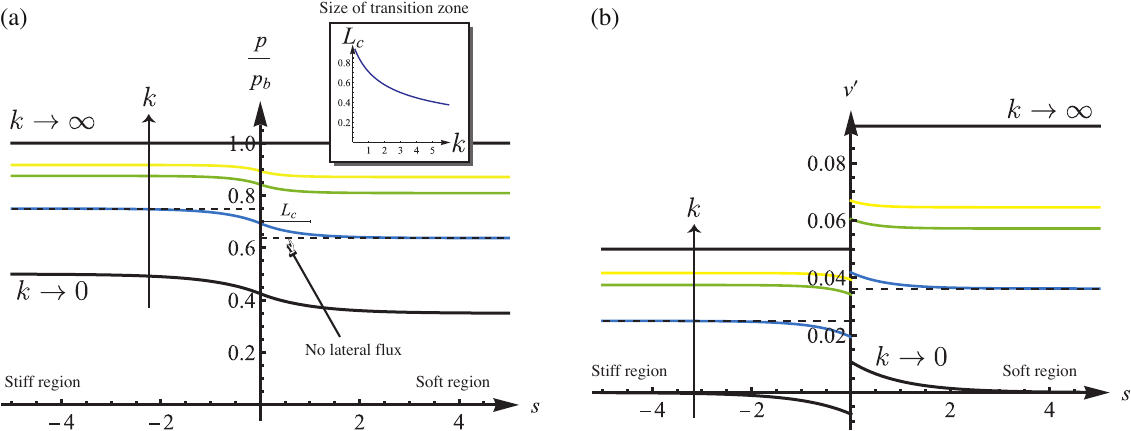}
    \caption{Asymptotic solution for small softening $\eta \ll 1$: (a) normalised pressure $p(s)/p^*$ and (b) expansion rate $v'(s)$, shown for different values of the effective  permeability with the outside $k$. Solid black lines show the limit cases $k\rightarrow 0$ (closed system) and $k\rightarrow 0$. Dashed lines show the case where flux is suppressed between the two regions. Inset shows the size of the transition zone $L_c$ as a function of $k$. Parameters: $\eta = 0.3$, $p^* - \epsilon^* = 0.05$, $k=1,3,5$.}
   \label{fig:asymptotic-inhibition}
\end{figure}

  Example solutions are shown in \cref{fig:asymptotic-inhibition}.  As can be seen, the softening of the right-hand-side region results in a pressure drop in that region, reflecting the reduced mechanical resistance of the cell walls to water entry. As pressure decreases smoothly at the junction between the soft to the stiff region,  the elongation rate $v'$ jumps discontinuously, reaching a global minimum at $s=0^-$ and a maximum at $s=0^+$. This jump directly results from the strain-based growth law, and the strain discontinuity at $s=0$. Due to the  pressure gradient, water seeps towards increasing $s$, i.e. from the stiff region to the soft region. The characteristic length of this seepage,  $L_c \eqdef -p'(0)/2\Delta p \leq 1$, increases when external water supply $k$ decreases, as illustrated by the inset in \cref{fig:asymptotic-inhibition}(a). This result shows the emergence of hydraulic competition between the two regions. This competition concerns a larger portion of the tissue when external water supply becomes small. The presence of an external source of water $k$ also tends to reduce the difference in pressure between the two regions (thus the seepage), as can be seen from the pressure drop $\Delta p = \eta p_0 / \pp{1+k}$ between the two asymptotes $s= -\infty$ and $s=+\infty$. This gap is maximal when $k\ll 1$, where water becomes scarce, with
   \begin{align}
     p(s) &=  \epsilon^* -  \frac\eta 2  \pp{\theta (-s)\E^s   + \theta (s) (2-\E^{-s})}+ O (\eta ^2).
   \end{align}
In this limit example however, the volume of water in the rod is conserved, thus as the right-hand side region grows along a region of effective size unity, the other side, which is actually under the growth threshold, has to shrink (thus this limit is actually nonphysical). This issue is addressed later in \cref{growth-induced-gradient}. For comparison, \cref{fig:asymptotic-inhibition} also shows the case where internal fluxes along the rod are suppressed ($K=0$),  illustrated by the dashed lines. In this case, the pressure and the velocity gradient are piece-wise continuous, with
\begin{equation}
    p(s)= p_0 - \frac{\eta p_0 \theta(s)}{k+1}+ O (\eta ^2),
\quad 
    v'(s) = a  + \frac{\eta  p_0\theta(s)}{1+k^{-1}} + O (\eta ^2),
\end{equation}
corresponding to the asymptotes for the general case. In the general case where the two regions exchange water, this jump in expansion rate $v'$ across the interface is amplified by a factor $ 1+k^{-1}$ with respect to the situation with zero flux. In particular, in situations where water supply is impeded (i.e. $k\ll 1$) and the overall growth becomes slower, this amplification factor diverges as the effect of hydraulic competition becomes more visible. In other words, although one might initially expect that water movements along the rod would serve to smooth out heterogeneities, we show that, near the interface ($\lvert s\rvert\lesssim L_c$), the combined effects of growth, hydraulics and mechanics actually accentuates the effect of heterogeneous mechanical properties on the growth dynamics. Interestingly, such effects have been proposed to play a role in the initiation of organs in the shoot apical meristem \cite{alonso2024water} (see \cref{discussion}).



\subsection{Water gradient in  apically growing organs \label{growth-induced-gradient}}

Plant shoots and roots grow typically along an elongating apical region \citep{erickson1976modeling}, while the rest of the tissue stiffens and stops growing. Kinematically, this type of growth is akin (but not physically equivalent) to tip  growth. Here, we explore the existence of tip-like, self-similar solutions to our system. Therefore, we focus on the steady regime \eqref{eqn:self-similar}, and define $s$, the arclength measured from the apex of the plant towards the base. In this scenario, $v$ is then the velocity of the domain measured in the co-moving frame attached to the apex.


To illustrate the role of chemical growth regulation in our framework, 
we here assume that growth is controlled by a growth hormone (e.g. auxin) with concentration $\rho$. We posit that the hormone is actively convected with velocity $v_a$ from the apex ($s=0$) towards the base ($s=\infty$), diffuses with diffusion coefficient $D$ and is reabsorbed with rate constant $\beta$. In the steady regime, the concentration profile obeys
\begin{equation}
D\rho'' - ( v+v_a)\rho' - (v'+\beta) \rho = 0 .
\end{equation}
In the convection-dominated regime $D \rho ', v\rho \ll v_a\rho$, the hormone concentration 
is given by 
$
    \rho (s) = \rho_0 \E^{-s/\sigma}$,
where $\rho_0 \eqdef \rho(0)$ is the concentration at the apex; and where $\sigma \eqdef \beta/v_a$ sets the length scale for the hormonal regulation  \citep{moulton2020multiscale}. In plants, transported hormones such as auxin are involved in the control of cell wall elastic properties \citep{sassi_auxin-mediated_2014}. Therefore, we assume that the effect of the hormone is to reduce the elastic modulus of the tissue. Assuming a small, linear effect of the hormone on the elastic modulus, we write 
\begin{equation}
    f\approx1 - \eta \rho/\rho_0,
\end{equation} 
with $\eta \ll 1$. Note that an alternative mechanism producing a gradient in rigidity  could be a gradual stiffening of the tissue as the cells age and move away from the apex \citep{eggen2011self}. 


Firstly, we assume that no growth  occurs in the absence of apical softening (i.e. if $\eta = 0$), thus we choose $\epsilon^*\geq p^*$. Indeed, in \cref{competition}, the choice $\epsilon^*<p^*$ resulted in exponential growth, precluding then the possibility of tip-like growth regimes. As previously, we solve \eqref{eqn:self-similar} asymptotically to first order in $\eta $. The base solution is $p_0=\epsilon_0=p^*$ and $v_0=0$. At $O (\eta )$, we have
\begin{equation}
    p_1'' -  (1+k) p_1   = p_0\E^{-s/\sigma}.\label{eqn:tip-growth-p1}
\end{equation}
Here $k$ can be interpreted biologically as a rate constant characterising the effective permeability between the growing tissue and the organ's vascular bundle.
Note that \eqref{eqn:tip-growth-p1} is valid only as long as the strain is above the growth threshold, i.e. $\epsilon>\epsilon^*$. Away from the tip, under this growth threshold, the system is in the purely elastic regime and obeys
\begin{equation}
    p_1'' -  k p_1   = 0.\label{eqn:tip-growth-p2}
\end{equation}
The general solutions for \eqref{eqn:tip-growth-p1,eqn:tip-growth-p2} can be obtained easily.
The difficulty is then to determine the unknown arclength $\Sigma$ at which the threshold is attained. Enforcing the condition that pressure $p$ is bounded and continuously differentiable, and using $\epsilon (\Sigma^-)=\epsilon (\Sigma^+) = \epsilon^*$ along with the no-flux condition at the tip $p'(0)=0$, we can express all the integration constants as functions of $\Sigma$. The latter is ultimately identified as the root of some (slightly uncomely) transcendental equation
which can be solved numerically, given  values of $k$, $\eta $, $\sigma$,  $\epsilon^*$, and $p^*$.

\begin{figure}[ht!]
    \centering
\includegraphics[width=\linewidth]{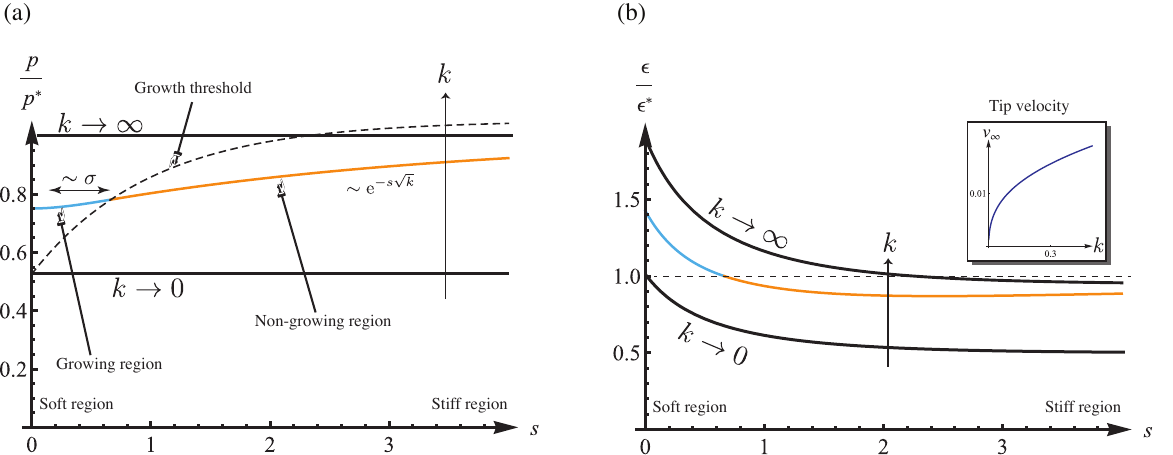}
    \caption{Asymptotic tip-like growth regimes for $\eta \ll 1$: (a) normalised pressure $p(s)/p^*$ and (b) normalised strain $\epsilon(s)/\epsilon^*$, shown for different values of $b$. The parts of the curves plotted in blue are above the growth threshold and correspond to growing regions; those in orange are under the threshold and correspond to a purely elastic regime. Solid black lines show the limit cases $k\rightarrow 0$ and $k\rightarrow \infty$. Dashed lines show the position of the growth threshold in pressure and strain space. Inset shows the tip velocity $v_\infty$ as a function of $k$. Parameters: $k=0.1$; $\epsilon^* = 0.1$; $p^*= 0.095$; $\sigma=1$; and $\eta = 0.5$.}
    \label{fig:tip-growth}
\end{figure}

\cref{fig:tip-growth} shows example solutions. As can be seen in \cref{fig:tip-growth}(a), the softening of the distal region results in a drop in pressure there. In the non-growing region (i.e. beyond the intersection point between the solid line and the dashed line), the pressure converges to its base value $p^*$ as $\E^{-s\sqrt{k}}$, i.e., perhaps surprisingly, there exists a water potential gradient of lengthscale $k^{-1/2}$ resulting from the softening of the shoot over a length  $\sigma$. 

We can compute the tip velocity with respect to the base as 
   $ v_\infty = v(\Sigma) $, which is, as expected, a growing function of $k$; see the inset in \cref{fig:tip-growth}(b). 
In the limit case $k\gg 1$, with $p=p^*$, the apex grows along a region of maximal size 
$\Sigma = \sigma \log \pp{ \eta\epsilon^* / \pp{\epsilon^*- p^*}}$. This length is positive only if $\eta > 1 - p^*/\epsilon^*$, i.e. when the equilibrium pressure $p^*$ is not too far from the growth threshold $\epsilon^*$, given a certain level of softening $\eta $, otherwise the pressure is too low to produce any growth at all. The tip velocity is then maximal, given  by
\begin{equation}
 v_\infty \approx  p^* \sigma  \pp{ {p^*}/{\epsilon^*}-1+\eta  }+\sigma  (p^*-\epsilon^*) \log \left(\frac{\eta  \epsilon^*}{\epsilon^*-p^*}\right).
\end{equation}
Conversely, when $k\rightarrow 0$, the growth zone vanishes and $v_\infty \rightarrow 0$. In this case, the base of the plant located at $s\rightarrow\infty$ is the only possible source of water but cannot sustain permanently the gradient in water potential necessary for growth. Both limit cases are plotted with solid black lines in \cref{fig:tip-growth}.

\subsection{Parameter estimates}

The biological relevance of hydraulic gradients is predicated upon the ratio of hydromechanical length $\mathcal L_h = \sqrt{KE\tau}$ to growth region size. To estimate this ratio, 
we consider a linear chain of identical Lockhart cells of average length $\ell $ (maintained constant through cell-division), hydraulically-insulated from their environment, but exchanging water with their two adjacent neighbours with membrane conductivity $\kappa$. From dimensional considerations, we expect the bulk conductivity for this chain of cells to be $K \sim \ell \kappa$. We take $\kappa = 10^{-8}$--$10^{-7} $\meter.\second\textsuperscript{-1}.\mega\pascal\textsuperscript{-1} \citep{laplaud2024assessing,forterre2022basic}. Taking the turgor pressure $p$ to be of the order of $1$ \mega\pascal{} \citep{long2020cellular} and the strain $\epsilon =0.05$, the bulk elastic modulus is $E=p/\epsilon=20$ \mega\pascal{}. Taking $\ell=10$ \micro\meter{} and estimating $\tau = 10^1$--$10^3$ \second{}, we obtain 
$\mathcal L_h  \approx 1$--$10 \ell$; namely, for this chain of cells, the length of interest for hydromechanical control is on the order of 1 to 10 cells. In practice, these values are very hard to measure with precision, and may vary a lot between different scenarios.







\section{Growth of a cylinder: hydraulics  and residual stress in  stem development\label{tube}}

 \subsection{Overview}
 
We now move on to a fully three-dimensional, nonlinearly elastic scenario to study the growth of a cylindrical stem. Here, the focus is on the interplay between heterogeneous material properties, water fluxes, and growth in multiple dimensions,  illustrating how hydraulic effects and differential material properties can constrain the growth and internal stresses within a simple three-dimensional structure.

 \subsection{Governing equations}

\begin{figure}[ht!]
    \centering
\includegraphics[width=.9\linewidth]{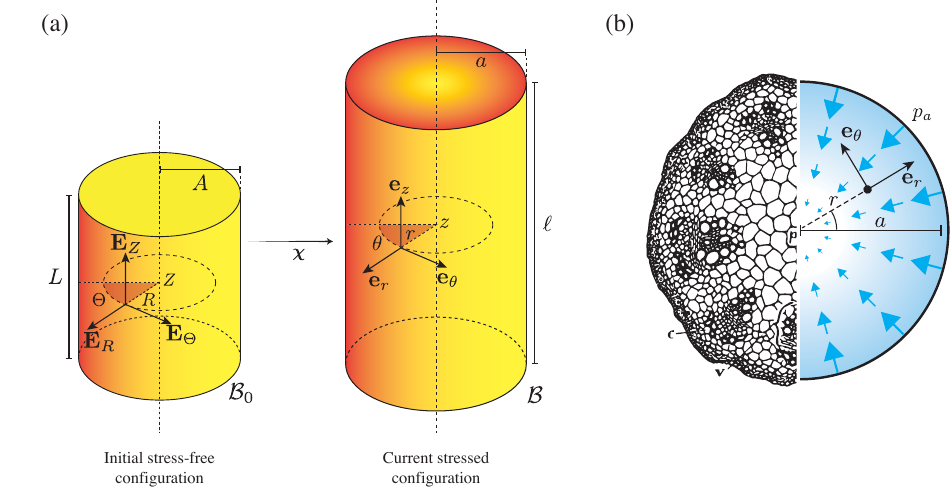}
    \caption{(a) Growth of a cylinder: initial and current configuration. (b) Hydraulic model for stem growth. Left: Cross-section of young stem of castor bean (\textit{Ricinus communis}): \textit{e} -- epidermis; \textit{c} -- cortex; \textit{p} -- 
pith; \textit{v} -- vascular bundle  \citep[drawing reproduced from][]{curtis1914nature}. Right: Schematic of the water fluxes in the cross-section. Water is supplied at the outer boundary, mimicking a vascular bundle located close to the epidermis.}
    \label{fig:cylinder-growth}
\end{figure}

We consider the growth of a cylinder of initial radius $A$ and length $L$, and current radius $a$ and length $\ell$. The initial domain is parameterised by the system of cylindrical coordinates  $(R,\Theta,Z)$, with $R$ the radial coordinate, $\Theta$ the azimuthal angle, and $Z$ the axial coordinate (\cref{fig:cylinder-growth}). In an axisymmetric deformation, a point $\pp{R,\Theta,Z}$ in the reference configuration is moved to the location $\pp{r,\theta,z}$ in the current configuration. Thus, the deformation $\boldsymbol{\chi}$ is given explicitly by
$
   r=  r(R,t)$, $ \theta =\Theta$, and $  z = z(Z,t) 
$, with 
\begin{equation}
    r(0,t) = 0, \quad z(0,t) = 0 \label{eqn:bc-r0}
\end{equation} 
at $R=0$.
In virtue of the symmetry, we can identify the two orthonormal bases associated with both systems of cylindrical coordinates: $\vec E_R=\vec e_r$, $\vec E_\Theta =\vec e_\theta$ and $\vec E_Z = \vec e_z$. 
Further assuming that the gradient of deformation is invariant by $Z$, the problem is effectively one-dimensional and depends only on $r$ and  $t$. The deformation gradient is given by
\begin{equation}
    \matr F = r'\, \vec e_r\otimes\vec E_R + \frac{r}{R} \,\vec e_\theta\otimes\vec E_\Theta +  \frac{\ell}{L} \,\vec e_z\otimes\vec E_Z,\label{eqn:deformation-gradient-polar}
\end{equation} 
where the apostrophe denotes differentiation w.r.t. $R$; and $\otimes$ is the tensor product. The Eulerian and Lagrangian velocities are given respectively by $\vec v  = v \,\vec e_r + z \dot\zeta \,\vec e_z $ and $\vec V  = V \,\vec e_r +(Z \dot \zeta \ell/L) \,\vec e_z$,  with $\dot\zeta \eqdef \dot\ell/\ell$ the relative rate of elongation.  The growth and elastic tensors can be written in the cylindrical basis as
  $ \matr G = \diag\pp{\gamma_r,\gamma_\theta,\gamma_z} $ and $\matr A = \diag\pp{\alpha_r,\alpha_\theta,\alpha_z}$. By \eqref{eqn:multiplicative,eqn:deformation-gradient-polar}, we have $r' = \alpha_r\gamma_r$, $r/R=\alpha_\theta\gamma_\theta$ and $\ell/L = \alpha_z\gamma_z$. The gradient of velocity can be related to the growth and elastic stretch rates through \eqref{eqn:velocity-gradient,eqn:rates}
\begin{subequations}\label{eqn:kinematic-relations-tube}
\begin{equation}
  \pdiff{v}{r} =\frac{ \dot \gamma_r}{ \gamma_r} + \frac{\dot\alpha_r}{\alpha_r} ,\quad 
 \frac{v}{r} = \frac{\dot \gamma_\theta}{\gamma_\theta} + \frac{\dot \alpha_\theta}{\alpha_\theta} ,
\quad 
    \dot \zeta = \frac{ \dot \gamma_z}{ \gamma_z} + \frac{\dot \alpha_z}{\alpha_z}.
\end{equation}
\end{subequations}

 By symmetry, the stress is also diagonal in the basis $\left\{\vec e_r, \vec e_\theta,\vec e_z \right\}$, with the total and partial Cauchy stresses given by $\matr T=\diag\pp{T_r,T_\theta,T_z}$ and $\matr T_s=\diag\pp{t_r,t_\theta,t_z}$, respectively. 
The only non-vanishing component in the balance of momentum \eqref{eqn:balance-linear-momentum2} is then
\begin{equation}
    \pdiff{t_r}{r} + \frac{t_r-t_{\theta} }{r} = \pdiff{p}{r}. 
    \label{eqn:linear-momentum-cylinder}
\end{equation}
Assuming that the end caps are free and not subject to any load, we obtain the boundary condition for the stresses at the end caps \citep[cf.][\S11.8.3]{Goriely2017},
\begin{equation}\label{eqn:tzintegral}
     \int_0^a (t_z - p) r\,\diff r= 0.
\end{equation}
Further, we assume that the outer pressure at $r=a$ is zero, so \eqref{eqn:bc-stress} yields the   condition
\begin{equation}
  p(a) = t_r(a)   .\label{eqn:ptrB}
\end{equation} 
The balance of mass \eqref{eqn:masss-final2} reads
\begin{equation}\label{balance-mass-tube}
     \frac{1}{r} \pdiff{}{r}\pp{r {v - K r \pdiff{\psi}{r} }} +  \dot \zeta   = \xi_f ,
\end{equation}
where $\matr K = K\matr 1$.
The source $\xi_f$ represents the water intake from a bulk vasculature  modelled as a single reservoir with water potential $\psi^*$, i.e. $\xi_f = k(\psi^* - \psi)$. Assuming that the system is far from hydraulic equilibrium, i.e. $\psi^* \gg \psi$, we have $\xi_f \approx  k\psi^*$, which is taken to be a function of $r$ only. Thus \eqref{balance-mass-tube} can be integrated directly as
\begin{equation}\label{eqn:integrated-balance-mass-tube}
          {v-K \pdiff{\psi}{r} } 
         =  \mathcal X_f - \frac{\dot\zeta  r}{2}   , 
         \quad \text{with} \quad 
    \mathcal X_f (r) \eqdef \frac{1}{r}\int_0^r \xi_f(r') r'\diff r'  ,
\end{equation}
where we have used the no-flux regularity condition $\linepdiff{\psi}{r} = 0$, and the geometric conditions $r=0$ and $v=0$ at $R=0$.
For simplicity, here we take $\xi_f$  constant so that $\mathcal X_f = \xi_f r /2$. Similarly the osmotic pressure $\pi$ is here assumed to not vary across the domain, so we elide the balance of osmolyte mass equation \eqref{eqn:balance-osmolyte-mass} and we take simply $\linepdiff{\psi}{r} = \linepdiff{p}{r}$. The cylinder exchanges water through its boundary at $R=A$, so we use the boundary condition \eqref{eqn:dirichlet} 
\begin{equation}
  \psi(a) =   \psi_a\quad \Leftrightarrow \quad p(a) = p_a, \label{eqn:fluxB2}
\end{equation}
with $p_a \eqdef \pi(a) + \psi_a$. In the example treated next, we assume $\xi_f=0$ so that water  enters the domain only through the boundary (however, for the sake of generality, we keep $\xi_f$ in the following derivations). In plants, the organisation of the vascular bundle may vary considerably between species.  The situation presented here mimics a type of architecture where the vascular bundle--xylem and phloem--is located near the epidermis, as illustrated in \cref{fig:cylinder-growth}(b). For more complex vascular systems, we may assume a non-zero source $\xi_f$, or an additional water exchange point at $R=0$ (as in roots, where the vascular bundle is located generally near the centre
). More in general, it is  easy to extend the problem to the case of a vascular bundle placed at an arbitrary position $ r_v \in\pp{0,a}$ by treating the two problems $r< r_v$ and $r> r_v$ separately \citep{passioura2003tissue}.

The growth law \eqref{eqn:strain-based-growth} corresponds to 
\begin{equation}
    \frac{\dot \gamma_i}{\gamma_i} =  \frac{1}{2\tau_i(R)} \ramp{\alpha_i^2-{\alpha^*}^2},\label{eqn:growth-laws-tube}
\end{equation}
with $i\in\left\{r,\theta,z\right\}$; and  $\tau_r$, $\tau_\theta$ and $\tau_z$ denote the characteristic times of material synthesis in the three separate directions of the cylinder. In growing stems, it is well known that heterogeneous material properties result in mechanical tension within the epidermis, a phenomenon called \textit{tissue tension}, manifesting the existence of residual stresses. These stresses emerge from differential growth of the tissue, where the core grows relatively faster than the epidermis   \citep{peters1996history,vandiver2008tissue,goriely2010elastic,holland2013mechanics,KellyBellow2023}. A classic experiment consists in peeling a stalk of rhubarb, which results in the detached cortex bending outward and shortening, along with rapid elongation of the exposed core when incubated in water \citep{sachs1865,kutschera1987cooperation,kutschera1989tissue,kutschera_epidermal-growth-control_2007,vandiver2008tissue,holland2013mechanics}, revealing tension in the cortex and compression below. This difference in growth rate is likely to be due to higher  growth extensibility of
the cell walls of the inner tissues \citep[as exposed in][and references therein]{kutschera1989tissue}. Therefore, we  assume that the characteristic time $\tau_z$ is larger near the epidermis than at the origin, so that, for equal strain levels, the epidermis undergoes slower growth. We posit  
$ \tau_z(R) = \tau_0 + \Delta \tau \pp{ {R}/{A}}^2$, where $\tau_0$ is the value at the origin, and  $\Delta \tau\geq 0$ defines the increment in $\tau_z$ between the core and the epidermis.  

Finally, the elastic constitutive equations read \begin{equation}\label{eqn:tube-constitutive-elastic}
    t_i = \frac{\alpha_i}{J_A} \pdiff{\Psi_s}{\alpha_i} .
\end{equation}
For simplicity, we here use an isotropic, compressible neo-Hookean strain energy function 
\begin{align}
    \Psi_s(\alpha_r, \alpha_\theta, \alpha_z) &=\frac{\mu}{2}    \left(\alpha_r^2+\alpha_\theta ^2+\alpha_z^2-3-2 \log (\alpha_r\alpha_\theta   \alpha_z)\right)  +\frac{\lambda }{2}  (\alpha_\theta  \alpha_r \alpha_z-1)^2,\label{eqn:neo-hookean}
\end{align}
where the material coefficients $\mu$ and $\lambda$ identify to the Lam\'e coefficients in the linear regime. Note that in the limit of incompressibility ($\lambda\gg 1$), no growth can occur since the solid cannot expand to absorb the fluid.  Henceforth we set $\lambda = \mu $ (corresponding to a Poisson ratio of 1/4 in the limit of linear elasticity). For simplicity, we assume that the elastic moduli are uniform \citep[see][and references therein]{kutschera1989tissue}. In fact, note that 
a heterogeneity in elastic rigidity would not be sufficient to capture differential growth on its own. Indeed, in a scenario where only $\mu$ would be spatially heterogeneous, the axial stretch $\alpha_z$ (thus the growth rate) would still be uniform in a cylindrical deformation.


As in \cref{rod}, we can nondimensionalise the system using $\sqrt{K\mu\tau_0}$, $\tau_0$ and $\mu$ as reference length, time and pressure, respectively. On eliminating $\dot\gamma_\theta/\gamma_\theta$ and $\dot\gamma_z/\gamma_z$ using \eqref{eqn:growth-laws-tube}, re-expressing the problem in the reference configuration, and rearranging the terms, we obtain a closed system of seven equations for the seven variables $\gamma_r$, $\alpha_r$, $\alpha_\theta$, $\alpha_z$, $V$, $r$, $p$, defined on the fixed domain $\mathcal B_0$:
\begin{subequations}
\label{eqn:nonhollow-tube-complete-system}
\begin{equation}
    r' =  \alpha_r\gamma_r \label{eqn:tube-rprime}
\end{equation}
    \begin{equation}
\frac{t_r ' - p'}{r'} =      \frac{t_\theta - t_r}{r} ,
    \label{eqn:linear-momentum-cylinder2}
\end{equation}
\begin{equation}\label{eqn:cylindre-balance-mass}
      \frac{p'}{r'} 
         = { \frac{1}{2}(\dot\zeta - \xi_f)   {r }  + V}    ,   
\end{equation}
\begin{equation}
    \frac{\dot \gamma_r}{\gamma_r} = \ramp{\alpha_r^2-{\alpha^*}^2},
\end{equation}
\begin{equation} 
     \frac{\dot \alpha_r}{\alpha_r} +  \frac{1}{2\tau_r }\ramp{\alpha_r^2-{\alpha^*}^2} = \frac{  V' }{r' }   ,
\end{equation}
\begin{equation}
 \frac{\dot \alpha_\theta}{\alpha_\theta} +   \frac{1}{2\tau_\theta }\ramp{\alpha_\theta^2-{\alpha^*}^2}= \frac{V}{r} ,\label{eqn:rate-theta}
\end{equation} 
\begin{equation} 
     \frac{\dot \alpha_z}{\alpha_z} +  \frac{1}{2\tau_z }\ramp{\alpha_z^2-{\alpha^*}^2} = \dot \zeta   ,
\end{equation}
\end{subequations}
where $t_r$ and $t_\theta$ are given in terms of the elastic stretches via \eqref{eqn:tube-constitutive-elastic}.
This system is equipped with the boundary conditions \eqref{eqn:bc-r0,eqn:fluxB2,eqn:ptrB} and the integral constraint \eqref{eqn:tzintegral} that is enforced via the undetermined parameter $\dot\zeta$. 
Note that \eqref{eqn:linear-momentum-cylinder2} has a geometric removable singularity at $R=0$ due to the boundary constraint $r(0,t)=0$, which generates computational difficulties. This issue is easily alleviated by considering a perturbed boundary condition $ r(\epsilon,t) \approx\epsilon \alpha_r(\epsilon,t) \gamma_r(\epsilon,t)$ at $R=\epsilon$, where $\epsilon\ll 1$. 






\subsection{Analysis of the solutions}

\subsubsection{Steady regime}

Before solving the full dynamical problem, we first restrict our attention to steady growth regimes for which we enforce $V=\dot \alpha_r=\dot \alpha_\theta=\dot\alpha_z=0$ and 
$\gamma_r=\gamma_\theta= 1$ (no radial growth). In this scenario, \eqref{eqn:cylindre-balance-mass} can be integrated directly, revealing that a permanent parabolic pressure profile is maintained across the elongating stem, as predicted by \cite{passioura2003tissue}:
\begin{equation}\label{eqn:steady-sol-stress}
    p - p_a = \frac{1}{4}(\dot\zeta - \xi_f) (r^2 - a^2).
\end{equation}
In particular, this pressure is non-negative if 
\begin{equation}\label{eqn:cylindre-inequality}
    a^2 \leq \frac{4p_a}{\dot\zeta - \xi_f};
\end{equation} 
otherwise, the radius is too large to allow for even distribution of water given the elongation rate $\dot \zeta$, and a zone of negative pressure forms at the centre of the stem. Eq.  \eqref{eqn:cylindre-inequality} may be viewed as a scaling constraint on growth, linking the  kinematics ($\dot \zeta$) and  geometry ($a$), for a given water supply ($p_a$, $\xi_f$).

\begin{figure}[ht!]
    \centering
\includegraphics[width=.9\linewidth]{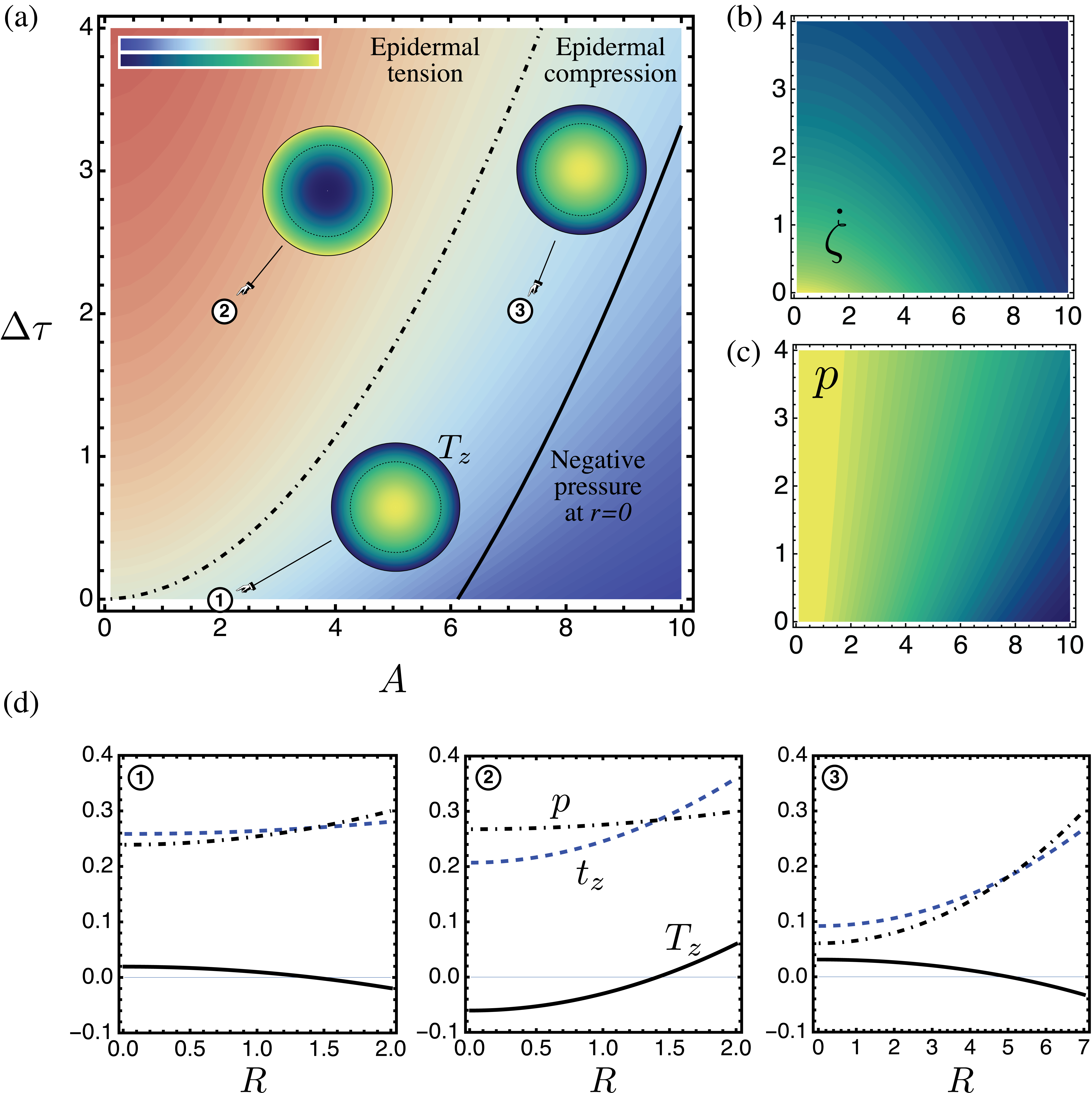}
    \caption{(a--c) Steady regime. (a) Contour plots shows differential tension $\Delta T_z = T_z (A) - T_z(0)$ between the outside and the centre of the cylinder, plotted vs. $A$ and  $\Delta \tau$. Solid line indicate the points where a zone of negative pressure appears at the centre of the stem. Dot dashed line show the level set $T_z(0) = T_z(A)$; above that line, the stress in the epidermis is higher than the stress at the centre. Insets show cross sections of the cylinder (scaled to the same radius) with colour indicating the axial stress $T_z$ increasing from blue to yellow, with dashed line indicating level sets $T_z=0$. (b) Rate of expansion $\dot \zeta$ vs. $A$ and $\Delta \tau$. (c) Mean pressure in the cross-section vs. $A$ and $\Delta \tau$. (d) Axial stress profiles and pressure in the various scenarios 1, 2 and 3 of (a). Black solid line shows the total stress $T_z=t_z-p$; Blue dashed line shows the partial stress $t_z$; black dot-dashed line shows turgor pressure $p$. Parameters: $p_a = 0.3$, $\lambda=1$, $\alpha^*=1$.}
    \label{fig:steady-cylinder}
\end{figure}

   The rest of the system can be solved numerically (\cref{numerics}). 
\cref{fig:steady-cylinder}(a--c) illustrate the steady solution and its dependency on the (nondimensionalised) stem radius $A$ and the parameter $\Delta \tau$. The density map in \cref{fig:steady-cylinder}(a) shows the difference in axial stress $T_z$ measured between the epidermis ($R=A$) and the origin ($R=0$), with red indicating states of epidermal tension where $T_z(A)>T_z(0)$, and dashed line showing the level set $T_z(A)=T_z(0)$. We show three cross-sections of the stem showing the stress distribution $T_z$ in three cases: 1--uniform $\tau_z$; 2--heterogeneous $\tau_z$ and thin stem; and 3--heterogeneous $\tau_z$ and thicker stem. \cref{fig:steady-cylinder}(b, c) show respectively the elongation rate $\dot \zeta$ and mean pressure, in the $A$-$\Delta \tau$ space. \cref{fig:steady-cylinder}(d) shows the profiles of pressure $p$ and stresses $T_z$ and $t_z$ for each state labelled 1, 2 or 3 in \cref{fig:steady-cylinder}(a).

For a uniform $\tau_z$, i.e. when $\Delta \tau=0$, the axial stress is compressive in the epidermis and tensile in the core, with $T_z(A)<0<T_z(0)$, see case 1 in \cref{fig:steady-cylinder}(a, d). This is purely a  hydraulic effect due to the higher pressure at the surface, $p(A)=p_a$. As can be seen, this heterogeneity in stress and pressure becomes more visible when the radius $A$ increases. When a gradient in wall extensibility is introduced, i.e. for  $\Delta \tau>0$, different scenarios are possible. For a relatively slender stem (i.e. above the  dashed line), a state of epidermal tissue tension is observed, where both the total stress $T_z$ and the partial solid stress $t_z$ are maximal at the epidermis, see case 2 in \cref{fig:steady-cylinder}(a, d). In this example, the core is in compression ($T_z(0)<0$); however, the solid matrix is generally still subject to tensile stresses ($t_z(R)>0$ for all $R$), indicating that the cells remain turgid and that their walls are indeed under tension, despite the overall compressive stress. This observation challenges the  conception that the cell walls  should be compressed and possibly buckled due to overall tissue compression \citep[as suggested in schematics by][]{kutschera1989tissue,peters1996history}. Overall, the distribution of stresses within a tissue is non trivial, in particular, the macroscopic stress--the one released upon cutting the tissue--is distinct from the stress experienced by the cell wall matrix. 

For thicker stems, below the dashed line in \cref{fig:steady-cylinder}(a), epidermal tension can no longer be maintained as hydraulic effects override the prescribed heterogeneity in extensibility. Indeed, as can be seen in \cref{fig:steady-cylinder}(c, d), the pressure becomes low near the origin, indicating a water deficit due to the increased distance to the source, as is especially visible in case 3 in \cref{fig:steady-cylinder}(a, d). As a result, a lower elongation rate $\dot \zeta$ is observed; see \cref{fig:steady-cylinder}(b). Here, even if the reduced epidermal extensibility would tend to promote epidermal tension, the epidermis is actually in compression, due to its better perfusion. A somewhat counterintuitive effect is observed where, unlike $T_z$, the axial stress $t_z$ actually increases with $R$, i.e. the tension in the solid matrix is maximal in the epidermis notwithstanding the global compression. This effect can be interpreted in light of the heterogeneous pressure profile: While the epidermis has high turgor pressure, generating tension in the walls but overall compression within the tissue (since the epidermis is constrained by the core), the pressure near the origin is too low to generate much tension in the cell walls, and most part of the solid stress there is provided by the  epidermis.

For even larger radii, i.e. below the solid line in \cref{fig:steady-cylinder}(a), a central region appears where pressure at the origin is negative, i.e. the inequality   \eqref{eqn:cylindre-inequality} is violated. This extreme growth-induced effect results from the high deficit in water, and from axial tensile forces applied to the core by the epidermis, which  effectively create a suction. While it is unclear whether such growth-induced negative pressure can exist, insofar as it results from the saturation assumption \eqref{eqn:saturation}, we suspect that the relative water deficit within the core, and the associated tensile stresses, could potentially participate in cavity opening during stem hollowing, described mathematically by \cite{goriely2010elastic}. 

\begin{figure}[ht!]
    \centering
    \includegraphics[width=0.4\linewidth]{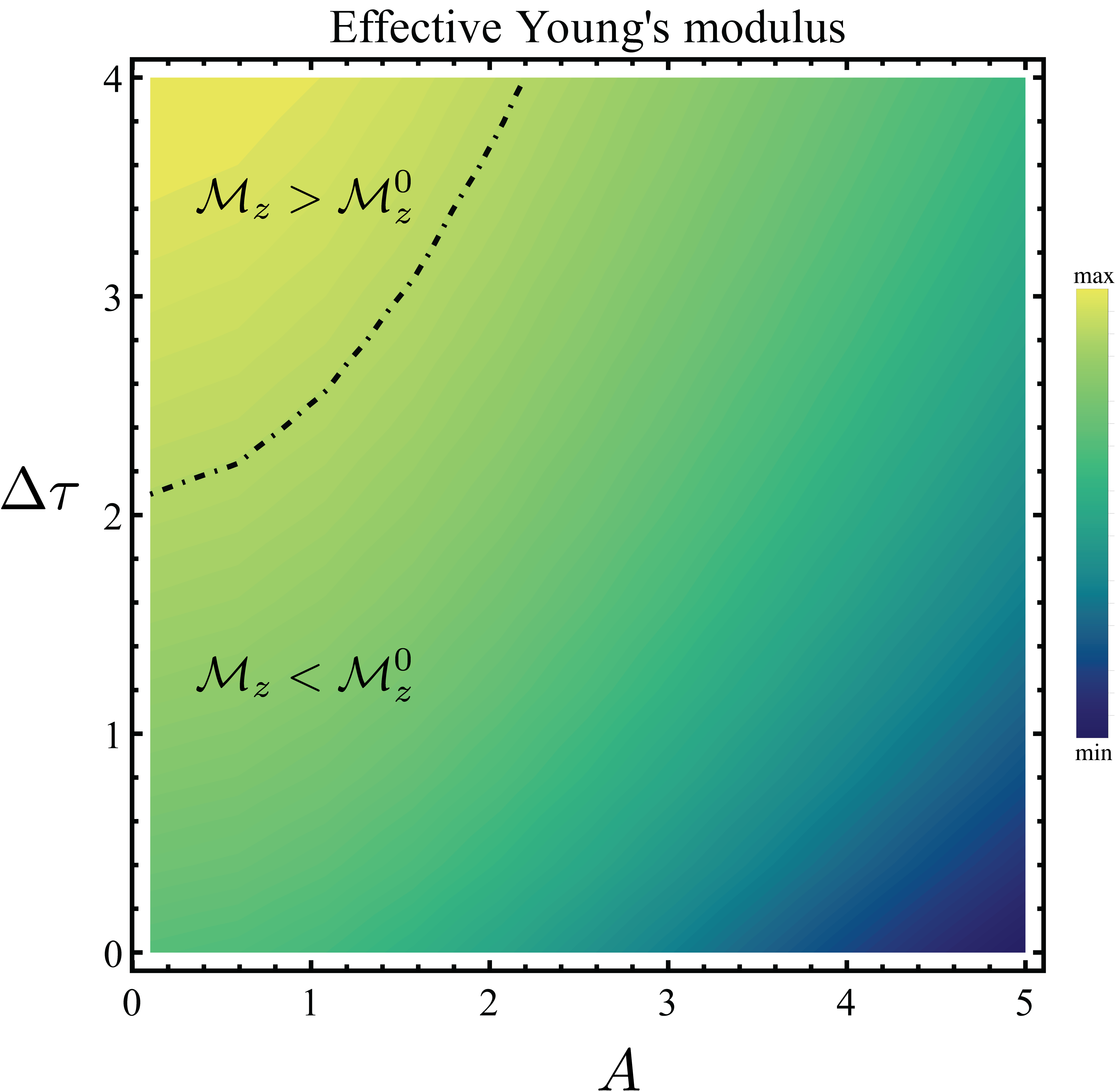}
    \caption{Apparent axial modulus $\mathcal M_z$ of the residually-stressed stem in $A$-$\Delta \tau$ space. Dashed line indicate the locations in the parameter space where $\mathcal M_z = \mathcal M_z^0$, i.e. where the axial modulus is that of the unstressed stem of identical radius.}
    \label{fig:stiffness}
\end{figure}

Following the developments of \cref{apparent-elasticity} we also assess the mechanical resistance of the stem under axial loads (i.e. its axial linear elastic modulus). We denote by $\hat{\matr A} = \diag( \hat \alpha_r , \hat\alpha_\theta, \hat\alpha_z)$ the incremental elastic deformation tensor, where $ \hat \alpha_r = \linepdiff{\hat r}{r}$ and $\hat \alpha_\theta = \hat r/r$, with $\hat r$ the radial coordinate in the incrementally deformed configuration.  
We first remark that the incompressibility condition
$
    (\hat \alpha_z{\hat r}/{r} ) \linepdiff{\hat r}{r} = 1
$
is separable and can be integrated directly as
$
  \hat r = r/  \sqrt{\hat\alpha_z}
$
from which we obtain 
\begin{equation}\label{eqn:incompressibility-cylinder}
    \hat\alpha_r = \hat\alpha_\theta = 1 /  \sqrt{\hat\alpha_z}.
\end{equation}
The stresses are given by \eqref{eqn:effective-stress} as
\begin{equation}\label{eqn:additive-stress-cylinder}
     \hat T_i =   t_i (\hat\alpha_r\alpha_r ,\hat\alpha_\theta\alpha_\theta ,\hat\alpha_z\alpha_z)- p  - \hat p ,
\end{equation}
with $i\in\left\{r,\theta,z\right\}$.
Taking the first variation of \eqref{eqn:incompressibility-cylinder,eqn:additive-stress-cylinder} around the base solution $\hat\alpha_i = 1$, we derive
\begin{equation}
     \delta \hat T_i = \mathcal T_i \,\delta\hat \alpha_z   - \delta \hat p, \quad\text{with}\quad  
    \mathcal T_i \eqdef  \pdiff{t_i}{\alpha_z} -   \frac12 \pp{\pdiff{t_i}{\alpha_r}+\pdiff{t_i}{\alpha_\theta}}.
\end{equation}
Then, on integrating the balance of momentum

\begin{equation}
   \pdiff{(\delta \hat T_r)}{ r} = \frac{\mathcal T_\theta - \mathcal T_r}{r}  \, \delta \hat \alpha_z,
\end{equation}
and using the identity $\delta \hat T_z - \delta \hat T_r= (\mathcal T_z   - \mathcal T_r) \,\delta \hat \alpha_z  $, we obtain finally the total virtual reaction force of the whole stem
\begin{equation}
    \delta F \eqdef 2\pi\int_0^a \delta T_z\,r\diff r = \mathcal M_z \delta \alpha_z ,  
\end{equation}
with $\mathcal M_z$ the effective longitudinal  linear elasticity modulus given by
\begin{align}\label{eqn:effective-stiffness}
    \mathcal M_z  =  2\pi  \int_0^a\pp{ \mathcal T_z   - \mathcal T_r -\int_r^a \frac{\mathcal T_\theta - \mathcal T_r}{r'}  \, \diff r'}r\diff r.
\end{align}
As can be seen, $\mathcal M_z$ depends on the pre-existing stresses and stretches within the stem.
In particular, in the absence of pre-stress ($\alpha_r=\alpha_\theta=\alpha_z=1$, $p=0$), we recover the known value $\mathcal M_z^0 = 3\pi  \mu B^2$ for the linear response of an incompressible neo-Hookean tube under uniaxial load. \cref{fig:stiffness} shows the dependency of the relative stiffness $\mathcal M_z / \mathcal M_z^0 $ on  $A$ and $\Delta \tau$. As can be seen by comparing the level sets of \cref{fig:stiffness} with those of \cref{fig:steady-cylinder}(a), the variation in axial stiffness of the stem is related to the presence of axial residual stresses, measured by $T_z(a) - T_z(0)$. This observation supports the hypothesis that tissue tension confers higher rigidity to the stem \citep{sachs1882vorlesungen,KUTSCHERA2001851}.
However, unfortunately, $\mathcal K_z$ does not inform us directly on the resistance of the stem to buckling, even for a thin stem, insofar as the residually-stressed cylinder is effectively heterogeneous and anisotropic. To that end, a full and likely tedious  perturbation analysis including asymmetric modes is required \citep{vandiver2008tissue,goriely2008nonlinear,moulton2020multiscale}. 

%


\subsubsection{Dynamic regime}
\begin{figure}[ht!]
    \centering
    \includegraphics[width=\linewidth]{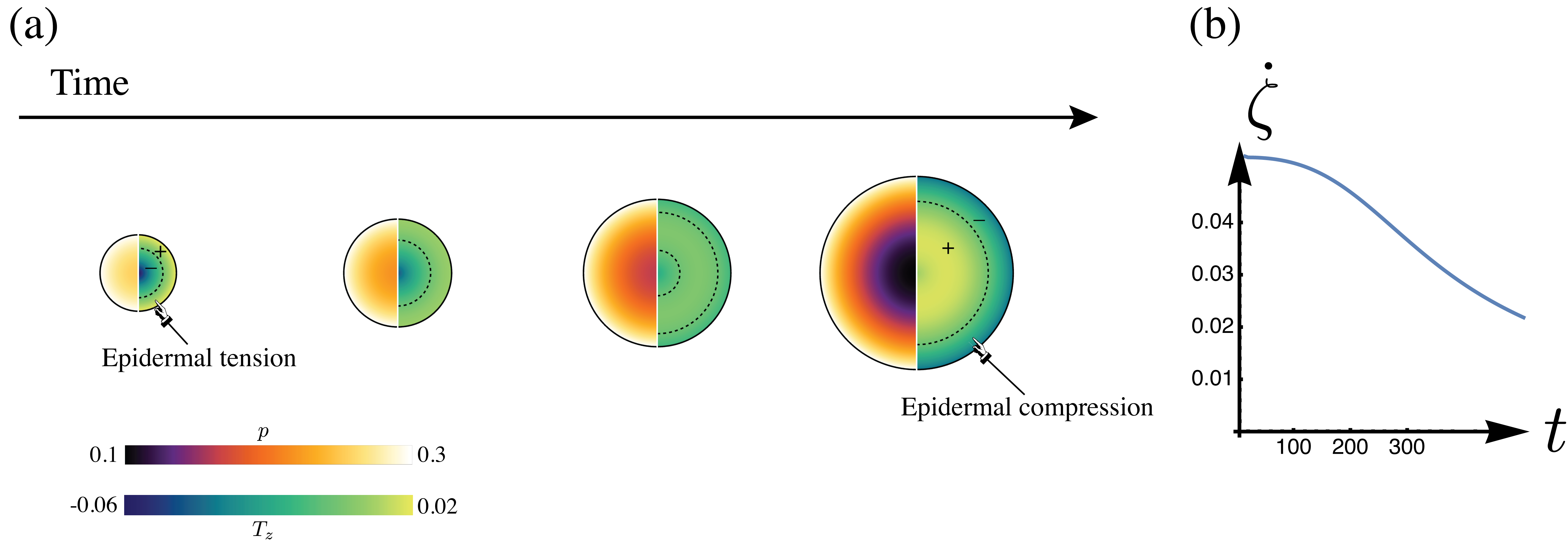}
    \caption{Dynamic regime. (a) Growing cross section of the stem for increasing time. The colour maps represent the turgor pressure $p$ on the left (from dark to light); and the tensile stress on the right (from blue to yellow) with level set $T_z=0$ shown with dashed line. Initially, the stem is in a state of epidermal tension due to positive $\Delta \tau$. Later on, pressure decreases in the core and epidermal tension vanishes. (b) Relative elongation rate $\dot\zeta$ vs. time. Parameters: $\Delta \tau=1$, $p_a = 0.3$, $\lambda=1$, $\alpha^*=1$, $A=1$.}
    \label{fig:dynamic-cylinder}
\end{figure}

These different states of the system can be observed in the full dynamic problem, as illustrated in \cref{fig:dynamic-cylinder}. Here, we simulate the three-dimensional growth of a stem with $\Delta \tau = 2$ taken as constant.  To account for the rapid elongation of the stem, we also assume that the extensibility is smaller in the $r$ and $\theta$ directions \citep[$\tau_r=\tau_\theta = 10 \tau_z$, consistant with values used by][]{KellyBellow2023}. As predicted earlier, epidermal tension is maintained until a critical radius is reached, at which point pressure becomes too low  at the origin and epidermal compression appears. As the radius increases and the mean pressure decreases, the elongation rate $\dot \zeta $ also decreases.

\section{Growth of a sphere\label{seed}}


As a final example, we study the full model \eqref{eqn:complete-set-equations} including the transport of osmolytes, that is, instead of assuming a constant osmotic pressure as in \cref{rod,tube}, we here assume that the osmolytes can diffuse on the growing domain, here a thick spherical shell. This example  illustrates how physiochemical details of organ growth can be included, for instance in the context of fruit growth which involves complex transport of sugar, gas exchanges and water fluxes. Although we do not aim here to model the sheer physiological complexity of fruit growth and maturation, our approach provides a paradigm to generalise detailed zero-dimensional multi-compartment approaches--e.g. \cite{cieslak2016integrating,Bussieres1994,fishman1998biophysical,martre2011modelling,dequeker2024biophysical}--to a continuum.

\begin{figure}[h!]
    \centering
    \includegraphics[width=\linewidth]{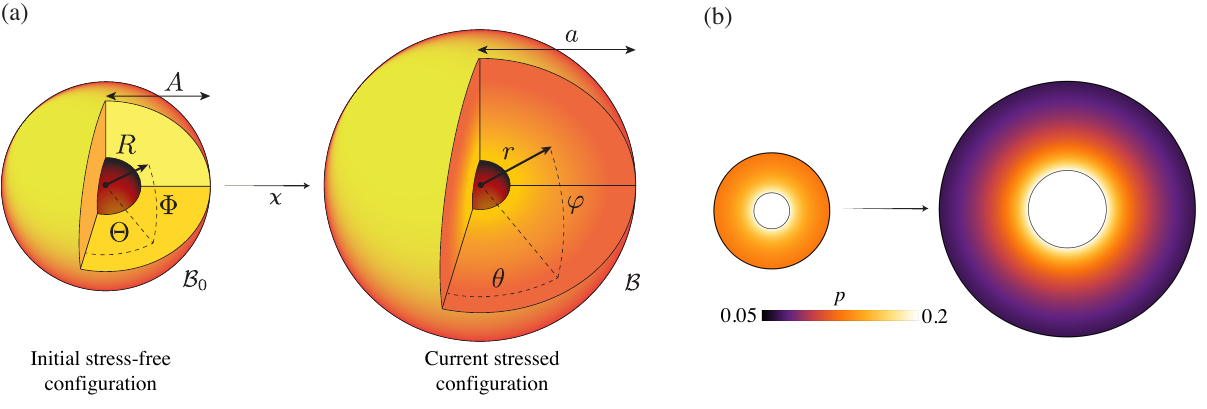}
    \caption{Growth of a hollow sphere. (a) Geometry: initial and current configuration with associated spherical coordinate systems. (b) Example simulation showing a drop in outer pressure during growth. Parameters: $A=0.03$, $B=0.1$, $\pi_a=0.2$, $k_b=0.5$, $\Psi_{ev}=0$, $\alpha^*=1$, $\lambda=1$, $D_{os}v_o = 1$.}
    \label{fig:sphere}
\end{figure}

We study the growth of a hollow sphere of inner and outer radius $a$ and $b$ in the current configuration ($A$ and $B$ respectively in the reference configuration). We introduce the system of spherical coordinates $(r,\theta,\varphi)$ in current configuration and  $(R,\Theta,\Phi)$ in reference configuration; see \cref{fig:sphere}(a). We assume the problem to be spherically symmetric so that, as in \cref{tube}, the problem only involves the coordinates $r$ and $t$.

For simplicity, we neglect active transport of the osmolytes ($
    \bar{\vec f}_o = 0
$) and we write $\matr L_{os} = (c_o R_g\theta / D_{os}) \matr 1$ with $D_{os}$ the diffusivity of the osmolytes in the solid, so that \eqref{eqn:high-drag-w} can be written as a standard Fickian flux
$
    \vec j_o =  - {D_{os}v_o } \linepdiff{c_o}{r}\, \vec e_r$.
We assume that the outer hydrostatic pressure is zero on both faces of the shell. For the flux of material, we assume a Dirichlet condition \eqref{eqn:dirichlet} at $r=a$. At the outer boundary $r=b$, we postulate an transpiration outflux associated with a water potential $\psi_{ev}$ and conductivity $k_b$. Altogether, we have
\begin{equation}\label{eqn:sphere-bc}
\begin{gathered}
       t_r(a,t) = p(a,t),\quad  \pi(a,t) = \pi_a; \quad \psi(a,t) =  \psi_a,\\  t_r(b,t) =p(b,t),\quad \pdiff{\psi}{r}(b,t)= k_b(\psi_{ev} - \psi),\quad \pdiff{\pi}{r}(b,t)=0,
\end{gathered}
\end{equation}
with $\psi_a$ the water potential across the inner boundary; and $\pi_a$ the osmotic pressure at $r=a$. We assume $\xi_f = \xi_o = 0$, so that the only source of water and osmolyte mass is the inner boundary. 
Following a procedure similar to that of \cref{tube}, we derive the (nondimensionalised) governing equations for the growing sphere:
\begin{subequations}
    \begin{equation}
    \pdiff{t_r}{r} + \frac{2}{r}\pp{t_r-t_\theta} = \pdiff{p}{r},
\end{equation}
\begin{equation}
          {v-\pdiff{}{r}(p-\pi ) }  = \frac{b^2}{r^2}\pp{ \dot b - k_b(\psi_{ev} - p_b + \pi_b) } , \label{eqn:balance-mass-sphere}
\end{equation}
\begin{equation}\label{eqn:balance-osmolyte-mass-sphere}
v\pi -  {  D_{os} v_o  \pdiff{\pi}{r} }   =         \frac{b^2   \dot b \pi_b  }{r^2}   ,
\end{equation}
\begin{equation}
     \frac{\dot \gamma_r}{\gamma_r} =  \frac12\ramp{\alpha_r^2 - {\alpha^*}^2},\quad 
    \frac{\dot \alpha_r}{\alpha_r} +  \frac12\ramp{\alpha_r^2 - {\alpha^*}^2} = \pdiff{v}{r},
\quad 
    \frac{\dot \alpha_\theta}{\alpha_\theta} +   \frac12\ramp{\alpha_\theta^2 - {\alpha^*}^2} = \frac{v}{r},
\end{equation}
\begin{equation}
    t_r = \frac{1}{\alpha_\theta^2}\pdiff{\Psi_s}{\alpha_r},\quad t_\theta = \frac{1}{\alpha_r\alpha_\theta}\pdiff{\Psi_s}{\alpha_\theta},
\end{equation}
\end{subequations}
where we have used the approximated van 't Hoff relation $\pi \approx R\theta\phi_o / v_o$, the two no-flux boundary conditions \eqref{eqn:sphere-bc} to integrate \eqref{eqn:masss-final2,eqn:masso-final2}, and the neo-Hookean strain energy \eqref{eqn:neo-hookean}; and where $\pi_b \eqdef \pi(b)$ and $p_b \eqdef p(b)$ are the undetermined osmotic and hydrostatic pressures at the outer boundary. We have assumed that the osmolyte diffusion is fast  and in the quasi-steady regime, so that we  set $\linepdiff{\phi_o}{t}=0$ in \eqref{eqn:masso-final2}. All parameters are taken to be uniform across the domain and constant in time. 
\cref{fig:sphere}(b) illustrates the evolution of pressure in a typical simulation. We see that during growth, the pressure drops in regions located far from the central source (similar to \cref{tube}). This simulation illustrates again the dynamic nature of pressure in the context of growth and osmotic regulations with multiple interfaces.
This approach may be beneficial to integrate spatiotemporal details of physiological and mechanical regulations in fruit development, as well as nonlocal couplings, which cannot be captured using more basic zero-dimensional models.

\section{Discussion\label{discussion}} 

While many continuum models of plant tissues morphogenesis have relied on phenomenological kinematic specifications of the  growth behaviour, the establishment of a mechanistic theory of morphogenesis requires modelling growth in relation to more fundamental physical and mechanical fields  \citep{Ambrosi2011,ambrosi2019growth,Goriely2017,menzel2012frontiers,kuhl2014growing}. 
In plants, it is widely accepted that cellular growth results from mechanical deformations of the cell walls driven by cell osmolarity. However, the explicit connection between pressure and growth has not been systematically integrated in continuum models. To bridge this gap, we proposed a formulation of plant developmental processes that extends the paradigm of Lockhart, Cosgrove and Ortega to a hydromechanical continuum description of the growth phenomenon. 


The role of pressure have been considered in numerous multicellular discrete models which have described growth as a result of turgor-induced cell wall expansion. However, typically, these models have treated pressure as a prescribed biological parameter, thereby neglecting the fundamentally mechanical nature of pressure. Physically, this modelling simplification originates in the assumption that  growth is primarily limited by the cell wall extensibility $\Phi$, thus, that water exchanges across cell membranes are instantaneous (i.e.  $p\approx\pi$). 
While this assumption may hold true in a first approximation, such simplification does not allow for a complete representation of the physical events occurring during growth. Recent reevaluation of water conductivity contributions in various systems has also revealed a  nuanced role of hydraulic effects, which may be non-negligible \citep{laplaud2024assessing,long2020cellular,alonso2024water}. These advancements motivate the development of new models based on a proper formulation of turgor. Indeed, the fundamental interplay between pressure, fluxes, osmolarity and cell mechanics is critical for a proper understanding of the notion of turgor, with profound experimental and conceptual implications. Thus, we assert that a robust physical description of growth should be grounded in clear balance relations. 
Our poroelastic theory  captures the growth of a tissue as the result of simultaneous mechanically-induced cell wall expansion and  water fluxes. Thus, pressure is treated as a dependent mechanical variable that  mediates  the  growth  deformations indirectly, through the balance of forces. This integrated perspective is directly in line with the original philosophy of Lockhart's approach.

A key mass balance principle can be stated as follows: For growth to occur, water must flow to fill the expanding volume, irrespective of the cause of cell wall expansion. In other words, growing regions correspond to hydraulic sinks and fast-growing regions are associated with lower water potential, as exemplified in \cref{rod}. Such sink is characterised by the development of growth-induced gradients of water potential and a water flux directed towards the growing region, an idea which has surfaced in the literature in particular in the work of Boyer and coworkers \citep{boyer1985control,boyer1988cell,nomami1993,passioura2003tissue,molz1978growth}. This flux, captured here through a Darcy-type law \eqref{eqn:darcy}, introduces a fundamental growth-induced \textit{hydromechanical length} $\mathcal L_h =\sqrt{KE\tau_0}$ which reflects the combined effects of growth ($\tau$), tissue permeability ($K$), and elasticity ($E$). For growing avascular domains larger than this typical length, interesting nonlocal couplings may emerge. For example, in \cref{competition}, we predicted that a fast growing region will absorb water from its neighbourhood, thereby hindering its growth. This idea is reminiscent of patterning mechanisms based on lateral inhibition \citep{Meinhardt2000}. Such phenomenon has been predicted in previous theoretical study \citep{cheddadi2019coupling} and was recently observed experimentally in \textit{Arabidopsis thaliana} shoot apices, where peripheral cells adjacent to incipient organs exhibit a shrinkage consistent with a water deficit \citep{alonso2024water}. Here we  characterise mathematically the magnitude and spatial extent of this inhibition (\cref{fig:asymptotic-inhibition}). In particular, we show that inhibition will be amplified and more spatially extended in  weakly vascularised tissues  ($k\ll 1$), where the inhibition zone is expected to have a width  $\sim \mathcal L_h$.

The role of hydromechanical effects is also illustrated in three dimensions in our model of a growing stem. Firstly, we showed that the classic epidermal tension hypothesis could be naturally captured by a gradient in cell extensibility; reproducing the phenomenology of previous continuum models \citep{KellyBellow2023,goriely2010elastic}. Secondly, for thick stems where the distance to the vascular bundle increases,  the distribution of residual stresses was greatly perturbed by hydraulic scaling effects when pressure at the core became very low. 
Naturally, in reality, the vascular architecture of plants may evolve as part of the developmental process to maintain adequate vascularisation. Therefore, we anticipate that our predictions may not apply universally, and the model must be adapted to capture more realistic scenarios. Overall, this work aims to exhibit guiding principles of hydromechanical control of plant morphogenesis and illustrate the complex, nonlocal and fundamentally dynamic behaviours that emerge from integrating fundamental growth mechanisms in space and time.

Several  questions remain open. The most pressing problem is to formulate a constitutive law describing growth, specifically the process by which the cell walls expand and solid mass is added to the system. 
 Although a strain-based growth law could capture the basic phenomenology of plant growth, its mechanistic interpretation is not fully established, in particular, it remains  unclear which specific mechanical quantity should be adopted as a driver of growth, and whether an appropriate growth law can be derived from rational mechanics considerations (\cref{growth-law}). Furthermore, a realistic growth law should also be able to capture the multiscale link between local cell structures and anisotropies and the overall growth of the tissue at the continuum level. In this context, promising efforts to derive continuum representations  from the cellular structures using multiscale analysis are emerging \citep{boudaoud2023multiscale}, and we hope that our work will motivate further advances in this area.
 

Another question concerns the link between the effective permeability tensor and the microscopic details of water routes within the tissue. In the context of direct cell-to-cell water transport  \citep{cheddadi2019coupling}, an approach would consist of  deriving the effective hydraulic conductivity of a periodically-repeating representative cell network  using two-scale analysis, as described by  
\cite{chapman2017effective}. More in general,  the biological details of water transport in different tissues are still an active subject of research, thus the precise biophysical interpretation of the effective conductivity is yet to be better characterised. For example, an interesting extension to our model would be to treat the apoplasmic route and the transmembrane exchanges between cell vacuoles separately \citep{molz1974water}. 

Lastly, a natural extension to this work is to model the role of  morphogens (e.g. hormones, such as auxin, or genes), using regular advection-diffusion equations, or more complex, nonlinear reaction-diffusion-advection-systems \citep{kierzkowski2019growth,NEWELL2008421,rueda2018curvature,kennaway2019volumetric,moulton2020multiscale,krause2023concentration}. In the spirit of our approach,  these morphogens should regulate specific physical and mechanical properties of the system, such as the rigidity, the osmolarity, the growth threshold or the cell extensibility, thereby indirectly influencing growth.

Overall, this work lays the foundations of a field theory of plant morphogenesis--a closed mathematical framework in which the growth phenomenon emerges as the product of multiple, coupled physical, chemical and mechanical fields acting more or less nonlocally. The construction of such theories is a formidable challenge in plant biomechanics and, in general, in the study of active living materials. 
In this broader context, the unassuming plant provides an interesting paradigm to build a general theory of living tissues.  


\addcontentsline{toc}{section}{Acknowledgements}
\section*{Acknowledgement}
 I.C. acknowledges support from the \textit{Institut rhônalpin des systèmes complexes (IXXI)}, and the \textit{Agence Nationale pour la Recherche} through the research project \textit{HydroField}. The authors are grateful to Alain Goriely, Andrea Giudici, Christophe Godin, and Arezki Boudaoud for insightful discussions.

\appendix

\section{Numerics and implementation\label{numerics}}
We integrate the system using a  relaxation method (backward time, centred space); see \cite{press2007numerical}. All simulations were implemented in the software \href{https://www.wolfram.com}{\textit{Wolfram Mathematica 14.0}}. Source code is available upon request.

\addcontentsline{toc}{section}{References}
\bibliographystyle{apalike}

\biboptions{sort&compress}

\begin{thebibliography}{}

\bibitem[Ali et~al., 2023]{ali2023revisiting}
Ali, O., Cheddadi, I., Landrein, B., and Long, Y. (2023).
\newblock Revisiting the relationship between turgor pressure and plant cell growth.
\newblock {\em New Phytologist}, 238(1):62--69.

\bibitem[Ali et~al., 2014]{ali2014physical}
Ali, O., Mirabet, V., Godin, C., and Traas, J. (2014).
\newblock Physical models of plant development.
\newblock {\em Annual Review of Cell and Developmental Biology}, 30(1):59--78.

\bibitem[Ali et~al., 2019]{ali2019simulating}
Ali, O., Oliveri, H., Traas, J., and Godin, C. (2019).
\newblock Simulating turgor-induced stress patterns in multilayered plant tissues.
\newblock {\em Bulletin of Mathematical Biology}, 81(8):3362--3384.

\bibitem[Ali and Traas, 2016]{ali_force-driven_2016}
Ali, O. and Traas, J. (2016).
\newblock Force-{Driven} {Polymerization} and {Turgor}-{Induced} {Wall} {Expansion}.
\newblock {\em Trends in Plant Science}, 21(5):398--409.

\bibitem[Alim et~al., 2012]{alim2012regulatory}
Alim, K., Hamant, O., and Boudaoud, A. (2012).
\newblock Regulatory role of cell division rules on tissue growth heterogeneity.
\newblock {\em Frontiers in Plant Science}, 3:174.

\bibitem[Alonso-Serra et~al., 2024]{alonso2024water}
Alonso-Serra, J., Cheddadi, I., Kiss, A., Cerutti, G., Lang, M., Dieudonn{\'e}, S., Lionnet, C., Godin, C., and Hamant, O. (2024).
\newblock Water fluxes pattern growth and identity in shoot meristems.
\newblock {\em Nature Communications}, 15(1):1--14.

\bibitem[Ambrosi et~al., 2011]{Ambrosi2011}
Ambrosi, D., Ateshian, G.~A., Arruda, E.~M., Cowin, S.~C., Dumais, J., Goriely, A., Holzapfel, G.~A., Humphrey, J.~D., Kemkemer, R., Kuhl, E., Olberding, J.~E., Taber, L.~A., and Garikipati, K.~R. (2011).
\newblock {Perspectives on biological growth and remodeling}.
\newblock {\em Journal of the Mechanics and Physics of Solids}, 59(4):863--883.

\bibitem[Ambrosi et~al., 2019]{ambrosi2019growth}
Ambrosi, D., Ben~Amar, M., Cyron, C.~J., DeSimone, A., Goriely, A., Humphrey, J.~D., and Kuhl, E. (2019).
\newblock Growth and remodelling of living tissues: perspectives, challenges and opportunities.
\newblock {\em Journal of the Royal Society Interface}, 16(157):20190233.

\bibitem[Ambrosi and Guana, 2007]{ambrosi2007stress}
Ambrosi, D. and Guana, F. (2007).
\newblock Stress-modulated growth.
\newblock {\em Mathematics and mechanics of solids}, 12(3):319--342.

\bibitem[Ambrosi et~al., 2012]{ambrosi2012interplay}
Ambrosi, D., Preziosi, L., and Vitale, G. (2012).
\newblock The interplay between stress and growth in solid tumors.
\newblock {\em Mechanics Research Communications}, 42:87--91.

\bibitem[Barbacci et~al., 2013]{barbacci2013another}
Barbacci, A., Lahaye, M., and Magnenet, V. (2013).
\newblock Another brick in the cell wall: biosynthesis dependent growth model.
\newblock {\em PLoS One}, 8(9):e74400.

\bibitem[Bassel et~al., 2014]{bassel2014mechanical}
Bassel, G.~W., Stamm, P., Mosca, G., Barbier~de Reuille, P., Gibbs, D.~J., Winter, R., Janka, A., Holdsworth, M.~J., and Smith, R.~S. (2014).
\newblock {Mechanical constraints imposed by 3D cellular geometry and arrangement modulate growth patterns in the {Arabidopsis} embryo}.
\newblock {\em Proceedings of the National Academy of Sciences}, page 201404616.

\bibitem[Bedford and Drumheller, 1983]{bedford1983theories}
Bedford, A. and Drumheller, D.~S. (1983).
\newblock Theories of immiscible and structured mixtures.
\newblock {\em International Journal of Engineering Science}, 21(8):863--960.

\bibitem[{Ben Amar} and Goriely, 2005]{BenAmar2005}
{Ben Amar}, M. and Goriely, A. (2005).
\newblock {Growth and instability in elastic tissues}.
\newblock {\em Journal of the Mechanics and Physics of Solids}, 53:2284–2319.

\bibitem[Bessonov et~al., 2013]{bessonov2013deformable}
Bessonov, N., Mironova, V., and Volpert, V. (2013).
\newblock Deformable cell model and its application to growth of plant meristem.
\newblock {\em Mathematical Modelling of Natural Phenomena}, 8(4):62--79.

\bibitem[Bou~Daher et~al., 2018]{daher2018anisotropic}
Bou~Daher, F., Chen, Y., Bozorg, B., Clough, J., J{\"o}nsson, H., and Braybrook, S.~A. (2018).
\newblock Anisotropic growth is achieved through the additive mechanical effect of material anisotropy and elastic asymmetry.
\newblock {\em eLife}, 7:e38161.

\bibitem[Boudaoud, 2010]{boudaoud2010introduction}
Boudaoud, A. (2010).
\newblock An introduction to the mechanics of morphogenesis for plant biologists.
\newblock {\em Trends in plant science}, 15(6):353--360.

\bibitem[Boudaoud et~al., 2023]{boudaoud2023multiscale}
Boudaoud, A., Kiss, A., and Ptashnyk, M. (2023).
\newblock Multiscale modeling and analysis of growth of plant tissues.
\newblock {\em SIAM Journal on Applied Mathematics}, 83(6):2354--2389.

\bibitem[Boudon et~al., 2015]{boudon_computational_2015}
Boudon, F., Chopard, J., Ali, O., Gilles, B., Hamant, O., Boudaoud, A., Traas, J., and Godin, C. (2015).
\newblock A {Computational} {Framework} for 3{D} {Mechanical} {Modeling} of {Plant} {Morphogenesis} with {Cellular} {Resolution}.
\newblock {\em PLoS Computational Biology Computational Biology}, 11(1):e1003950.

\bibitem[Boyer, 1988]{boyer1988cell}
Boyer, J.~S. (1988).
\newblock Cell enlargement and growth-induced water potentials.
\newblock {\em Physiologia Plantarum}, 73(2):311--316.

\bibitem[Boyer et~al., 1985]{boyer1985control}
Boyer, J.~S., Cavalieri, A., and Schulze, E.~D. (1985).
\newblock Control of the rate of cell enlargement: excision, wall relaxation, and growth-induced water potentials.
\newblock {\em Planta}, 163:527--543.

\bibitem[Bozorg et~al., 2016]{bozorg_continuous_2016}
Bozorg, B., Krupinski, P., and J\"onsson, H. (2016).
\newblock A continuous growth model for plant tissue.
\newblock {\em Physical Biology}, 13(6):065002.

\bibitem[Bussières, 1994]{Bussieres1994}
Bussières, P. (1994).
\newblock {Water Import Rate in Tomato Fruit: A Resistance Model}.
\newblock {\em Annals of Botany}, 73(1):75--82.

\bibitem[Chakraborty et~al., 2021]{CHAKRABORTY2021110736}
Chakraborty, J., Luo, J., and Dyson, R.~J. (2021).
\newblock Lockhart with a twist: Modelling cellulose microfibril deposition and reorientation reveals twisting plant cell growth mechanisms.
\newblock {\em Journal of Theoretical Biology}, 525:110736.

\bibitem[Chapman and Shabala, 2017]{chapman2017effective}
Chapman, S.~J. and Shabala, A. (2017).
\newblock Effective transport properties of lattices.
\newblock {\em SIAM Journal on Applied Mathematics}, 77(5):1631--1652.

\bibitem[Cheddadi et~al., 2019]{cheddadi2019coupling}
Cheddadi, I., G{\'e}nard, M., Bertin, N., and Godin, C. (2019).
\newblock Coupling water fluxes with cell wall mechanics in a multicellular model of plant development.
\newblock {\em PLoS Computational Biology}, 15(6):e1007121.

\bibitem[Chickarmane et~al., 2010]{chickarmane2010computational}
Chickarmane, V., Roeder, A.~H., Tarr, P.~T., Cunha, A., Tobin, C., and Meyerowitz, E.~M. (2010).
\newblock Computational morphodynamics: a modeling framework to understand plant growth.
\newblock {\em Annual review of plant biology}, 61:65--87.

\bibitem[Cieslak et~al., 2016]{cieslak2016integrating}
Cieslak, M., Cheddadi, I., Boudon, F., Baldazzi, V., G{\'e}nard, M., Godin, C., and Bertin, N. (2016).
\newblock Integrating physiology and architecture in models of fruit expansion.
\newblock {\em Frontiers in plant science}, 7:1739.

\bibitem[Coen, 1999]{coen2000art}
Coen, E. (1999).
\newblock {\em The art of genes: How organisms make themselves}.
\newblock Oxford University Press, Oxford.

\bibitem[Coen and Cosgrove, 2023]{coen2023mechanics}
Coen, E. and Cosgrove, D.~J. (2023).
\newblock The mechanics of plant morphogenesis.
\newblock {\em Science}, 379(6631):eade8055.

\bibitem[Coen et~al., 2017]{coen2017genes}
Coen, E., Kennaway, R., and Whitewoods, C. (2017).
\newblock On genes and form.
\newblock {\em Development}, 144(23):4203--4213.

\bibitem[Coen et~al., 2004]{coen_genetics_2004}
Coen, E., Rolland-Lagan, A.-G., Matthews, M., Bangham, J.~A., and Prusinkiewicz, P. (2004).
\newblock The genetics of geometry.
\newblock {\em Proceedings of the National Academy of Sciences of the United States of America}, 101(14):4728--4735.

\bibitem[Coleman and Noll, 1963]{Coleman1963167}
Coleman, B.~D. and Noll, W. (1963).
\newblock The thermodynamics of elastic materials with heat conduction and viscosity.
\newblock {\em Archive for Rational Mechanics and Analysis}, 13(1):167 – 178.

\bibitem[Corson et~al., 2009]{corson_turning_2009}
Corson, F., Hamant, O., Bohn, S., Traas, J., Boudaoud, A., and Couder, Y. (2009).
\newblock Turning a plant tissue into a living cell froth through isotropic growth.
\newblock {\em Proceedings of the National Academy of Sciences}, 106(21):8453--8458.

\bibitem[Cosgrove, 1981]{cosgrove1981}
Cosgrove, D.~J. (1981).
\newblock Analysis of the dynamic and steady-state responses of growth rate and turgor pressure to changes in cell parameters.
\newblock {\em Plant Physiology}, 68(6):1439--1446.

\bibitem[Cosgrove, 1985]{cosgrove1985cell}
Cosgrove, D.~J. (1985).
\newblock Cell wall yield properties of growing tissue: evaluation by in vivo stress relaxation.
\newblock {\em Plant physiology}, 78(2):347--356.

\bibitem[Cosgrove, 1993]{cosgrove1993water}
Cosgrove, D.~J. (1993).
\newblock Water uptake by growing cells: an assessment of the controlling roles of wall relaxation, solute uptake, and hydraulic conductance.
\newblock {\em International journal of plant sciences}, 154(1):10--21.

\bibitem[Cosgrove, 2005]{cosgrove2005growth}
Cosgrove, D.~J. (2005).
\newblock Growth of the plant cell wall.
\newblock {\em Nature reviews molecular cell biology}, 6(11):850--861.

\bibitem[Cosgrove, 2016]{cosgrove2016plant}
Cosgrove, D.~J. (2016).
\newblock Plant cell wall extensibility: connecting plant cell growth with cell wall structure, mechanics, and the action of wall-modifying enzymes.
\newblock {\em Journal of experimental botany}, 67(2):463--476.

\bibitem[Cosgrove, 2018]{cosgrove2018diffuse}
Cosgrove, D.~J. (2018).
\newblock Diffuse growth of plant cell walls.
\newblock {\em Plant physiology}, 176(1):16--27.

\bibitem[Cosgrove and Anderson, 2020]{cosgrove2020plant}
Cosgrove, D.~J. and Anderson, C.~T. (2020).
\newblock Plant cell growth: do pectins drive lobe formation in {Arabidopsis} pavement cells?
\newblock {\em Current Biology}, 30(11):R660--R662.

\bibitem[Coussy, 2003]{coussy2004poromechanics}
Coussy, O. (2003).
\newblock {\em Poromechanics}.
\newblock John Wiley \& Sons, Chichester.

\bibitem[Curtis, 1914]{curtis1914nature}
Curtis, C.~C. (1914).
\newblock {\em Nature and development of plants}.
\newblock Henry Holt and Company, New York, 4 edition.

\bibitem[Dainty, 1963]{DAINTY1963279}
Dainty, J. (1963).
\newblock Water relations of plant cells.
\newblock In Preston, R.~D., editor, {\em Volume 1}, Advances in Botanical Research, pages 279--326. Academic Press.

\bibitem[De~Boer, 1992]{de1992development}
De~Boer, R. (1992).
\newblock Development of porous media theories—a brief historical review.
\newblock {\em Transport in porous media}, 9:155--164.

\bibitem[De~Boer, 2012]{de2012theory}
De~Boer, R. (2012).
\newblock {\em Theory of porous media: highlights in historical development and current state}.
\newblock Springer Berlin, Heidelberg.

\bibitem[Dequeker et~al., 2024]{dequeker2024biophysical}
Dequeker, B., {\v{S}}alagovi{\v{c}}, J., Retta, M., Verboven, P., and Nicola{\"\i}, B.~M. (2024).
\newblock {A biophysical model of apple (Malus domestica Borkh.) and pear (Pyrus communis L.) fruit growth}.
\newblock {\em Biosystems Engineering}, 239:130--146.

\bibitem[Dervaux and Ben~Amar, 2008]{dervaux2008morphogenesis}
Dervaux, J. and Ben~Amar, M. (2008).
\newblock Morphogenesis of growing soft tissues.
\newblock {\em Physical Review Letters}, 101:068101.

\bibitem[DiCarlo and Quiligotti, 2002]{DICARLO2002449}
DiCarlo, A. and Quiligotti, S. (2002).
\newblock Growth and balance.
\newblock {\em Mechanics Research Communications}, 29(6):449--456.

\bibitem[Dumais, 2021]{dumais2021}
Dumais, J. (2021).
\newblock Mechanics and hydraulics of pollen tube growth.
\newblock {\em New Phytologist}, 232(4):1549--1565.

\bibitem[Dumais and Forterre, 2012]{dumais2012vegetable}
Dumais, J. and Forterre, Y. (2012).
\newblock {“Vegetable dynamicks”: the role of water in plant movements}.
\newblock {\em Annual Review of Fluid Mechanics}, 44:453--478.

\bibitem[Dunlop et~al., 2010]{dunlop2010theoretical}
Dunlop, J.~W., Fischer, F.~D., Gamsj{\"a}ger, E., and Fratzl, P. (2010).
\newblock A theoretical model for tissue growth in confined geometries.
\newblock {\em Journal of the Mechanics and Physics of Solids}, 58(8):1073--1087.

\bibitem[Dupuy et~al., 2007]{dupuy2007system}
Dupuy, L., Mackenzie, J., Rudge, T., and Haseloff, J. (2007).
\newblock A system for modelling cell-cell interactions during plant morphogenesis.
\newblock {\em Annals of botany}, 101(8):1255--1265.

\bibitem[Dyson et~al., 2012]{dyson_model_2012}
Dyson, R.~J., Band, L.~R., and Jensen, O.~E. (2012).
\newblock A model of crosslink kinetics in the expanding plant cell wall: {Yield} stress and enzyme action.
\newblock {\em Journal of Theoretical Biology}, 307:125--136.

\bibitem[Eggen et~al., 2011]{eggen2011self}
Eggen, E., de~Keijzer, M.~N., and Mulder, B.~M. (2011).
\newblock Self-regulation in tip-growth: The role of cell wall ageing.
\newblock {\em Journal of theoretical biology}, 283(1):113--121.

\bibitem[Epstein and Maugin, 2000]{epstein2000thermomechanics}
Epstein, M. and Maugin, G.~A. (2000).
\newblock Thermomechanics of volumetric growth in uniform bodies.
\newblock {\em International Journal of Plasticity}, 16(7-8):951--978.

\bibitem[Erickson, 1976]{erickson1976modeling}
Erickson, R.~O. (1976).
\newblock Modeling of plant growth.
\newblock {\em Annual review of plant physiology}, 27(1):407--434.

\bibitem[Fishman and G{\'e}nard, 1998]{fishman1998biophysical}
Fishman, S. and G{\'e}nard, M. (1998).
\newblock A biophysical model of fruit growth: simulation of seasonal and diurnal dynamics of mass.
\newblock {\em Plant, Cell \& Environment}, 21(8):739--752.

\bibitem[Forterre, 2013]{forterre2013slow}
Forterre, Y. (2013).
\newblock Slow, fast and furious: understanding the physics of plant movements.
\newblock {\em Journal of experimental botany}, 64(15):4745--4760.

\bibitem[Forterre, 2022]{forterre2022basic}
Forterre, Y. (2022).
\newblock Basic soft matter for plants.
\newblock In Jensen, K. and Forterre, Y., editors, {\em Soft Matter in Plants: From Biophysics to Biomimetics}, number~15 in Soft Matter Series, chapter~1, pages 1--65. The Royal Society of Chemistry, London.

\bibitem[Fozard et~al., 2013]{fozard_vertex-element_2013}
Fozard, J.~A., Lucas, M., King, J.~R., and Jensen, O.~E. (2013).
\newblock Vertex-element models for anisotropic growth of elongated plant organs.
\newblock {\em Frontiers in Plant Science}, 4(233).

\bibitem[Fraldi and Carotenuto, 2018]{fraldi2018cells}
Fraldi, M. and Carotenuto, A.~R. (2018).
\newblock Cells competition in tumor growth poroelasticity.
\newblock {\em Journal of the Mechanics and Physics of Solids}, 112:345--367.

\bibitem[Fridman et~al., 2021]{fridman2021root}
Fridman, Y., Strauss, S., Horev, G., Ackerman-Lavert, M., Reiner-Benaim, A., Lane, B., Smith, R., and Savaldi-Goldstein, S. (2021).
\newblock The root meristem is shaped by brassinosteroid control of cell geometry.
\newblock {\em Nature plants}, 7(11):1475--1484.

\bibitem[Geitmann and Ortega, 2009]{geitmann_mechanics_2009}
Geitmann, A. and Ortega, J. K.~E. (2009).
\newblock Mechanics and modeling of plant cell growth.
\newblock {\em Trends in Plant Science}, 14(9):467--478.

\bibitem[Goriely, 2017]{Goriely2017}
Goriely, A. (2017).
\newblock {\em {The mathematics and mechanics of biological growth}}, volume~45 of {\em Interdisciplinary applied mathematics}.
\newblock Springer-Verlag, New York.

\bibitem[Goriely et~al., 2010]{goriely2010elastic}
Goriely, A., Moulton, D.~E., and Vandiver, R. (2010).
\newblock Elastic cavitation, tube hollowing, and differential growth in plants and biological tissues.
\newblock {\em Europhysics Letters}, 91(1):18001.

\bibitem[Goriely et~al., 2008a]{goriely_elastic_2008}
Goriely, A., Robertson-Tessi, M., Tabor, M., and Vandiver, R. (2008a).
\newblock Elastic growth models.
\newblock In Mondaini, R.~P. and Pardalos, P.~M., editors, {\em Mathematical modelling of biosystems}, volume 102 of {\em Applied Optimization}, chapter~1, pages 1--44. Springer-Verlag Berlin Heidelberg.

\bibitem[Goriely and Tabor, 1998]{goriely1998spontaneous}
Goriely, A. and Tabor, M. (1998).
\newblock Spontaneous helix hand reversal and tendril perversion in climbing plants.
\newblock {\em Physical Review Letters}, 80:1564--1567.

\bibitem[Goriely et~al., 2008b]{goriely2008nonlinear}
Goriely, A., Vandiver, R., and Destrade, M. (2008b).
\newblock {Nonlinear Euler buckling}.
\newblock {\em Proceedings of the Royal Society A: Mathematical, Physical and Engineering Sciences}, 464(2099):3003--3019.

\bibitem[Green et~al., 2010]{green_genetic_2010}
Green, A.~A., Kennaway, J.~R., Hanna, A.~I., Bangham, J.~A., and Coen, E. (2010).
\newblock Genetic {Control} of {Organ} {Shape} and {Tissue} {Polarity}.
\newblock {\em PLoS Computational Biology Biol}, 8(11):e1000537.

\bibitem[Gurtin et~al., 2010]{gurtin2010mechanics}
Gurtin, M.~E., Fried, E., and Anand, L. (2010).
\newblock {\em The mechanics and thermodynamics of continua}.
\newblock Cambridge University Press, New York.

\bibitem[Haas et~al., 2020]{haas2020pectin}
Haas, K.~T., Wightman, R., Meyerowitz, E.~M., and Peaucelle, A. (2020).
\newblock Pectin homogalacturonan nanofilament expansion drives morphogenesis in plant epidermal cells.
\newblock {\em Science}, 367(6481):1003--1007.

\bibitem[Haas et~al., 2021]{HAAS2021100054}
Haas, K.~T., Wightman, R., Peaucelle, A., and Höfte, H. (2021).
\newblock The role of pectin phase separation in plant cell wall assembly and growth.
\newblock {\em The Cell Surface}, 7:100054.

\bibitem[Hamant et~al., 2008]{hamant2008developmental}
Hamant, O., Heisler, M.~G., Jonsson, H., Krupinski, P., Uyttewaal, M., Bokov, P., Corson, F., Sahlin, P., Boudaoud, A., Meyerowitz, E.~M., Couder, Y., and Traas, J. (2008).
\newblock Developmental patterning by mechanical signals in {Arabidopsis}.
\newblock {\em Science}, 322(5908):1650--1655.

\bibitem[Hamant and Traas, 2010]{hamant2010mechanics}
Hamant, O. and Traas, J. (2010).
\newblock The mechanics behind plant development.
\newblock {\em New Phytologist}, 185(2):369--385.

\bibitem[Holland et~al., 2013]{holland2013mechanics}
Holland, M.~A., Kosmata, T., Goriely, A., and Kuhl, E. (2013).
\newblock On the mechanics of thin films and growing surfaces.
\newblock {\em Mathematics and Mechanics of Solids}, 18(6):561--575.

\bibitem[Holzapfel, 2000]{holzapfel2000nonlinear}
Holzapfel, G.~A. (2000).
\newblock {\em Nonlinear solid mechanics}.
\newblock John Wiley \& Sons, Chichester.

\bibitem[Jia et~al., 2018]{jia2018curvature}
Jia, F., Pearce, S.~P., and Goriely, A. (2018).
\newblock Curvature delays growth-induced wrinkling.
\newblock {\em Physical Review E}, 98(3):033003.

\bibitem[Johnson and Lenhard, 2011]{johnson2011genetic}
Johnson, K. and Lenhard, M. (2011).
\newblock Genetic control of plant organ growth.
\newblock {\em New Phytologist}, 191(2):319--333.

\bibitem[Kelly-Bellow et~al., 2023]{KellyBellow2023}
Kelly-Bellow, R., Lee, K., Kennaway, R., Barclay, J.~E., Whibley, A., Bushell, C., Spooner, J., Yu, M., Brett, P., Kular, B., Cheng, S., Chu, J., Xu, T., Lane, B., Fitzsimons, J., Xue, Y., Smith, R.~S., Whitewoods, C.~D., and Coen, E. (2023).
\newblock Brassinosteroid coordinates cell layer interactions in plants via cell wall and tissue mechanics.
\newblock {\em Science}, 380(6651):1275--1281.

\bibitem[Kennaway and Coen, 2019]{kennaway2019volumetric}
Kennaway, R. and Coen, E. (2019).
\newblock Volumetric finite-element modelling of biological growth.
\newblock {\em Open biology}, 9(5):190057.

\bibitem[Kennaway et~al., 2011]{kennaway_generation_2011}
Kennaway, R., Coen, E., Green, A., and Bangham, A. (2011).
\newblock Generation of {Diverse} {Biological} {Forms} through {Combinatorial} {Interactions} between {Tissue} {Polarity} and {Growth}.
\newblock {\em PLoS Computational Biology}, 7(6):e1002071.

\bibitem[Khadka et~al., 2019]{khadka2019feedback}
Khadka, J., Julien, J.-D., and Alim, K. (2019).
\newblock Feedback from tissue mechanics self-organizes efficient outgrowth of plant organ.
\newblock {\em Biophysical journal}, 117(10):1995--2004.

\bibitem[Kierzkowski et~al., 2012]{kierzkowski_elastic_2012}
Kierzkowski, D., Nakayama, N., Routier-Kierzkowska, A.-L., Weber, A., Bayer, E., Schorderet, M., Reinhardt, D., Kuhlemeier, C., and Smith, R.~S. (2012).
\newblock Elastic domains regulate growth and organogenesis in the plant shoot apical meristem.
\newblock {\em Science}, 335(6072):1096--1099.

\bibitem[Kierzkowski et~al., 2019]{kierzkowski2019growth}
Kierzkowski, D., Runions, A., Vuolo, F., Strauss, S., Lymbouridou, R., Routier-Kierzkowska, A.-L., Wilson-S{\'a}nchez, D., Jenke, H., Galinha, C., Mosca, G., et~al. (2019).
\newblock A growth-based framework for leaf shape development and diversity.
\newblock {\em Cell}, 177(6):1405--1418.

\bibitem[Krause et~al., 2023]{krause2023concentration}
Krause, A.~L., Gaffney, E.~A., and Walker, B.~J. (2023).
\newblock Concentration-dependent domain evolution in reaction--diffusion systems.
\newblock {\em Bulletin of Mathematical Biology}, 85(2):14.

\bibitem[Kuhl, 2014]{kuhl2014growing}
Kuhl, E. (2014).
\newblock Growing matter: A review of growth in living systems.
\newblock {\em Journal of the Mechanical Behavior of Biomedical Materials}, 29:529--543.

\bibitem[Kutschera, 1989]{kutschera1989tissue}
Kutschera, U. (1989).
\newblock Tissue stresses in growing plant organs.
\newblock {\em Physiologia Plantarum}, 77(1):157--163.

\bibitem[Kutschera, 2001]{KUTSCHERA2001851}
Kutschera, U. (2001).
\newblock Gravitropism of axial organs in multicellular plants.
\newblock {\em Advances in Space Research}, 27(5):851--860.

\bibitem[Kutschera et~al., 1987]{kutschera1987cooperation}
Kutschera, U., Bergfeld, R., and Schopfer, P. (1987).
\newblock Cooperation of epidermis and inner tissues in auxin-mediated growth of maize coleoptiles.
\newblock {\em Planta}, 170(2):168--180.

\bibitem[Kutschera and Niklas, 2007]{kutschera_epidermal-growth-control_2007}
Kutschera, U. and Niklas, K.~J. (2007).
\newblock The epidermal-growth-control theory of stem elongation: An old and a new perspective.
\newblock {\em Journal of Plant Physiology}, 164(11):1395--1409.

\bibitem[Lang et~al., 2015]{lang2015propagation}
Lang, G.~E., Vella, D., Waters, S.~L., and Goriely, A. (2015).
\newblock Propagation of damage in brain tissue: coupling the mechanics of oedema and oxygen delivery.
\newblock {\em Biomechanics and modeling in mechanobiology}, 14:1197--1216.

\bibitem[Laplaud et~al., 2024]{laplaud2024assessing}
Laplaud, V., Muller, E., Demidova, N., Drevensek, S., and Boudaoud, A. (2024).
\newblock Assessing the hydromechanical control of plant growth.
\newblock {\em Journal of the Royal Society Interface}, 21(214):20240008.

\bibitem[Lee et~al., 2019]{lee2019shaping}
Lee, K. J.~I., Bushell, C., Koide, Y., Fozard, J.~A., Piao, C., Yu, M., Newman, J., Whitewoods, C., Avondo, J., Kennaway, R., Marée, A. F.~M., Cui, M., and Coen, E. (2019).
\newblock Shaping of a three-dimensional carnivorous trap through modulation of a planar growth mechanism.
\newblock {\em PLoS Biology}, 17(10):e3000427.

\bibitem[Liang and Mahadevan, 2009]{liang2009shape}
Liang, H. and Mahadevan, L. (2009).
\newblock The shape of a long leaf.
\newblock {\em Proceedings of the National Academy of Sciences}, 106(52):22049--22054.

\bibitem[Liu et~al., 2022]{LIU20221974}
Liu, S., Strauss, S., Adibi, M., Mosca, G., Yoshida, S., {Dello Ioio}, R., Runions, A., Andersen, T.~G., Grossmann, G., Huijser, P., Smith, R.~S., and Tsiantis, M. (2022).
\newblock Cytokinin promotes growth cessation in the {Arabidopsis} root.
\newblock {\em Current Biology}, 32(9):1974--1985.e3.

\bibitem[Liu et~al., 2013]{liu2013pattern}
Liu, Z., Swaddiwudhipong, S., and Hong, W. (2013).
\newblock Pattern formation in plants via instability theory of hydrogels.
\newblock {\em Soft Matter}, 9(2):577--587.

\bibitem[Lockhart, 1965]{lockhart1965analysis}
Lockhart, J.~A. (1965).
\newblock An analysis of irreversible plant cell elongation.
\newblock {\em Journal of theoretical biology}, 8(2):264--275.

\bibitem[Long et~al., 2020]{long2020cellular}
Long, Y., Cheddadi, I., Mosca, G., Mirabet, V., Dumond, M., Kiss, A., Traas, J., Godin, C., and Boudaoud, A. (2020).
\newblock Cellular heterogeneity in pressure and growth emerges from tissue topology and geometry.
\newblock {\em Current Biology}, 30(8):1504--1516.

\bibitem[Martre et~al., 2011]{martre2011modelling}
Martre, P., Bertin, N., Salon, C., and G{\'e}nard, M. (2011).
\newblock Modelling the size and composition of fruit, grain and seed by process-based simulation models.
\newblock {\em New Phytologist}, 191(3):601--618.

\bibitem[Martyushev and Seleznev, 2006]{martyushev2006maximum}
Martyushev, L.~M. and Seleznev, V.~D. (2006).
\newblock Maximum entropy production principle in physics, chemistry and biology.
\newblock {\em Physics reports}, 426(1):1--45.

\bibitem[Meinhardt and Gierer, 2000]{Meinhardt2000}
Meinhardt, H. and Gierer, A. (2000).
\newblock {Pattern formation by local self-activation and lateral inhibition}.
\newblock {\em BioEssays}, 22(8):753--760.

\bibitem[Menzel and Kuhl, 2012]{menzel2012frontiers}
Menzel, A. and Kuhl, E. (2012).
\newblock Frontiers in growth and remodeling.
\newblock {\em Mechanics research communications}, 42:1--14.

\bibitem[Merks et~al., 2011]{merks2011virtualleaf}
Merks, R.~M., Guravage, M., Inz{\'e}, D., and Beemster, G.~T. (2011).
\newblock Virtual{L}eaf: an open-source framework for cell-based modeling of plant tissue growth and development.
\newblock {\em Plant physiology}, 155(2):656--666.

\bibitem[Molz and Boyer, 1978]{molz1978growth}
Molz, F.~J. and Boyer, J.~S. (1978).
\newblock Growth-induced water potentials in plant cells and tissues.
\newblock {\em Plant Physiology}, 62(3):423--429.

\bibitem[Molz and Ikenberry, 1974]{molz1974water}
Molz, F.~J. and Ikenberry, E. (1974).
\newblock Water transport through plant cells and cell walls: theoretical development.
\newblock {\em Soil Science Society of America Journal}, 38(5):699--704.

\bibitem[Molz et~al., 1975]{molz1975dynamics}
Molz, F.~J., Truelove, B., and Peterson, C.~M. (1975).
\newblock Dynamics of rehydration in leaf disks 1.
\newblock {\em Agronomy Journal}, 67(4):511--515.

\bibitem[Mosca et~al., 2018]{mosca2018modeling}
Mosca, G., Adibi, M., Strauss, S., Runions, A., Sapala, A., and Smith, R.~S. (2018).
\newblock Modeling plant tissue growth and cell division.
\newblock In Morris, R.~J., editor, {\em Mathematical modelling in plant biology}, pages 107--138. Springer, Cham.

\bibitem[Mosca et~al., 2024]{mosca2024growth}
Mosca, G., Eng, R.~C., Adibi, M., Yoshida, S., Lane, B., Bergheim, L., Weber, G., Smith, R.~S., and Hay, A. (2024).
\newblock Growth and tension in explosive fruit.
\newblock {\em Current Biology}, 34:1010–1022.

\bibitem[Moulton et~al., 2020]{moulton2020multiscale}
Moulton, D.~E., Oliveri, H., and Goriely, A. (2020).
\newblock Multiscale integration of environmental stimuli in plant tropism produces complex behaviors.
\newblock {\em Proceedings of the National Academy of Sciences}, 117(51):32226--32237.

\bibitem[Newell et~al., 2008]{NEWELL2008421}
Newell, A.~C., Shipman, P.~D., and Sun, Z. (2008).
\newblock Phyllotaxis: Cooperation and competition between mechanical and biochemical processes.
\newblock {\em Journal of Theoretical Biology}, 251(3):421--439.

\bibitem[Nonami and Boyer, 1993]{nomami1993}
Nonami, H. and Boyer, J.~S. (1993).
\newblock Direct demonstration of a growth-induced water potential gradient.
\newblock {\em Plant Physiology}, 102(1):13--19.

\bibitem[Oliveri et~al., 2024]{Oliveri2024}
Oliveri, H., Moulton, D.~E., Harrington, H.~A., and Goriely, A. (2024).
\newblock Active shape control by plants in dynamic environments.
\newblock {\em Physical Review E}, 110(1):014405.

\bibitem[Oliveri et~al., 2018]{oliveri2019regulation}
Oliveri, H., Traas, J., Godin, C., and Ali, O. (2018).
\newblock Regulation of plant cell wall stiffness by mechanical stress: a mesoscale physical model.
\newblock {\em Journal of mathematical biology}, 78(3):625--653.

\bibitem[Onsager, 1931]{onsager1931reciprocal}
Onsager, L. (1931).
\newblock {Reciprocal relations in irreversible processes. I.}
\newblock {\em Physical review}, 37(4):405.

\bibitem[Ortega, 1985]{ortega_augmented_1985}
Ortega, J. K.~E. (1985).
\newblock Augmented {Growth} {Equation} for {Cell} {Wall} {Expansion}.
\newblock {\em Plant Physiology}, 79(1):318--320.

\bibitem[Ortega, 2010]{ortega2010plant}
Ortega, J. K.~E. (2010).
\newblock Plant cell growth in tissue.
\newblock {\em Plant physiology}, 154(3):1244--1253.

\bibitem[Passioura and Boyer, 2003]{passioura2003tissue}
Passioura, J.~B. and Boyer, J.~S. (2003).
\newblock Tissue stresses and resistance to water flow conspire to uncouple the water potential of the epidermis from that of the xylem in elongating plant stems.
\newblock {\em Functional Plant Biology}, 30(3):325--334.

\bibitem[Peng et~al., 2022]{peng2022differential}
Peng, Z., Alique, D., Xiong, Y., Hu, J., Cao, X., L{\"u}, S., Long, M., Wang, Y., Wabnik, K., and Jiao, Y. (2022).
\newblock Differential growth dynamics control aerial organ geometry.
\newblock {\em Current Biology}, 32(22):4854--4868.

\bibitem[Peters and Tomos, 1996]{peters1996history}
Peters, W.~S. and Tomos, A.~D. (1996).
\newblock The history of tissue tension.
\newblock {\em Annals of Botany}, 77(6):657--665.

\bibitem[Philip, 1958]{philip1958propagation}
Philip, J. (1958).
\newblock Propagation of turgor and other properties through cell aggregations.
\newblock {\em Plant Physiology}, 33(4):271.

\bibitem[Pieczywek and Zdunek, 2017]{pieczywek2017compression}
Pieczywek, P.~M. and Zdunek, A. (2017).
\newblock {Compression simulations of plant tissue in 3D using a mass-spring system approach and discrete element method}.
\newblock {\em Soft matter}, 13(40):7318--7331.

\bibitem[Plant, 1982]{plant1982continuum}
Plant, R.~E. (1982).
\newblock A continuum model for root growth.
\newblock {\em Journal of Theoretical Biology}, 98(1):45--59.

\bibitem[Press et~al., 2007]{press2007numerical}
Press, W.~H., Teukolsky, S.~A., Vetterling, W.~T., and Flannery, B.~P. (2007).
\newblock {\em Numerical recipes: The art of scientific computing}.
\newblock Cambridge University Press, New York, 3 edition.

\bibitem[Preziosi and Farina, 2002]{preziosi2002darcy}
Preziosi, L. and Farina, A. (2002).
\newblock {On Darcy's law for growing porous media}.
\newblock {\em International Journal of Non-Linear Mechanics}, 37(3):485--491.

\bibitem[Rebocho et~al., 2017]{Rebocho2017}
Rebocho, A.~B., Southam, P., Kennaway, J.~R., Bangham, J.~A., and Coen, E. (2017).
\newblock Generation of shape complexity through tissue conflict resolution.
\newblock {\em eLife}, 6:e20156.

\bibitem[Robinson and Kuhlemeier, 2018]{robinson2018global}
Robinson, S. and Kuhlemeier, C. (2018).
\newblock Global compression reorients cortical microtubules in {A}rabidopsis hypocotyl epidermis and promotes growth.
\newblock {\em Current Biology}.

\bibitem[Rodriguez et~al., 1994]{rodriguez1994stress}
Rodriguez, E.~K., Hoger, A., and McCulloch, A.~D. (1994).
\newblock {Stress-dependent finite growth in soft elastic tissues}.
\newblock {\em Journal of Biomechanics}, 27(4):455--467.

\bibitem[Rojas et~al., 2011]{rojas_chemically_2011}
Rojas, E.~R., Hotton, S., and Dumais, J. (2011).
\newblock {Chemically {Mediated} {Mechanical} {Expansion} of the {Pollen} {Tube} {Cell} {Wall}}.
\newblock {\em Biophysical Journal}, 101(8):1844--1853.

\bibitem[Rudge and Haseloff, 2005]{rudge2005computational}
Rudge, T. and Haseloff, J. (2005).
\newblock A computational model of cellular morphogenesis in plants.
\newblock In {\em European Conference on Artificial Life}, pages 78--87. Springer.

\bibitem[Rueda-Contreras et~al., 2018]{rueda2018curvature}
Rueda-Contreras, M.~D., Romero-Arias, J.~R., Aragon, J.~L., and Barrio, R.~A. (2018).
\newblock {Curvature-driven spatial patterns in growing 3D domains: A mechanochemical model for phyllotaxis}.
\newblock {\em PLoS One}, 13(8):e0201746.

\bibitem[{Sachs}, 1865]{sachs1865}
{Sachs}, J. (1865).
\newblock {\em Handbuch der Experimental-Physiologie der Pflanzen: Untersuchungen über die allgemeinen Lebensbedingungen der Pflanzen und die Functionen ihrer Organe}, volume~4 of {\em Handbuch der Experimental-physiologie der Pflanzen}.
\newblock Wilhelm Engelmann, Leipzig.

\bibitem[Sachs, 1882]{sachs1882vorlesungen}
Sachs, J. (1882).
\newblock {\em Vorlesungen {\"u}ber Pflanzen-physiologie}, volume~1.
\newblock W. Engelmann, Leipzig.

\bibitem[Sassi et~al., 2014]{sassi_auxin-mediated_2014}
Sassi, M., Ali, O., Boudon, F., Cloarec, G., Abad, U., Cellier, C., Chen, X., Gilles, B., Milani, P., Friml, J., Vernoux, T., Godin, C., Hamant, O., and Traas, J. (2014).
\newblock An {Auxin}-{Mediated} {Shift} toward {Growth} {Isotropy} {Promotes} {Organ} {Formation} at the {Shoot} {Meristem} in {Arabidopsis}.
\newblock {\em Current Biology}, 24(19):2335--2342.

\bibitem[Schopfer, 2006]{schopfer2006biomechanics}
Schopfer, P. (2006).
\newblock Biomechanics of plant growth.
\newblock {\em American journal of botany}, 93(10):1415--1425.

\bibitem[Tiero and Tomassetti, 2016]{tiero2016morphoelastic}
Tiero, A. and Tomassetti, G. (2016).
\newblock On morphoelastic rods.
\newblock {\em Mathematics and Mechanics of Solids}, 21(8):941--965.

\bibitem[Vandiver and Goriely, 2008]{vandiver2008tissue}
Vandiver, R. and Goriely, A. (2008).
\newblock Tissue tension and axial growth of cylindrical structures in plants and elastic tissues.
\newblock {\em Europhysics Letters}, 84(5):58004.

\bibitem[Vandiver and Goriely, 2009]{vandiver2009morpho}
Vandiver, R. and Goriely, A. (2009).
\newblock Morpho-elastodynamics: the long-time dynamics of elastic growth.
\newblock {\em Journal of biological dynamics}, 3(2-3):180--195.

\bibitem[Wada, 2012]{wada2012hierarchical}
Wada, H. (2012).
\newblock Hierarchical helical order in the twisted growth of plant organs.
\newblock {\em Physical Review Letters}, 109(12):128104.

\bibitem[Wang and Zhao, 2015]{wang2015three}
Wang, Q. and Zhao, X. (2015).
\newblock A three-dimensional phase diagram of growth-induced surface instabilities.
\newblock {\em Scientific reports}, 5(1):8887.

\bibitem[Whitewoods and Coen, 2017]{whitewoods2017growth}
Whitewoods, C.~D. and Coen, E. (2017).
\newblock Growth and development of three-dimensional plant form.
\newblock {\em Current Biology}, 27(17):R910--R918.

\bibitem[Whitewoods et~al., 2020]{whitewoods2020evolution}
Whitewoods, C.~D., Gon{\c{c}}alves, B., Cheng, J., Cui, M., Kennaway, R., Lee, K., Bushell, C., Yu, M., Piao, C., and Coen, E. (2020).
\newblock Evolution of carnivorous traps from planar leaves through simple shifts in gene expression.
\newblock {\em Science}, 367(6473):91--96.

\bibitem[Xue et~al., 2016]{XUE2016409}
Xue, S.-L., Li, B., Feng, X.-Q., and Gao, H. (2016).
\newblock Biochemomechanical poroelastic theory of avascular tumor growth.
\newblock {\em Journal of the Mechanics and Physics of Solids}, 94:409--432.

\bibitem[Zhang et~al., 2024]{zhang2024mechanism}
Zhang, H., Xue, F., Guo, L., Cheng, J., Jabbour, F., DuPasquier, P.-E., Xie, Y., Zhang, P., Wu, Y., Duan, X., Kong, H., and Zhang, R. (2024).
\newblock {The mechanism underlying asymmetric bending of lateral petals in \textit{Delphinium} (Ranunculaceae)}.
\newblock {\em Current Biology}, 34:755–768.

\bibitem[Zhang et~al., 2020]{zhang2020wox}
Zhang, Z., Runions, A., Mentink, R.~A., Kierzkowski, D., Karady, M., Hashemi, B., Huijser, P., Strauss, S., Gan, X., Ljung, K., and Tsiantis, M. (2020).
\newblock {A WOX/auxin biosynthesis module controls growth to shape leaf form}.
\newblock {\em Current biology}, 30(24):4857--4868.

\bibitem[Zhao et~al., 2020]{zhao2020microtubule}
Zhao, F., Du, F., Oliveri, H., Zhou, L., Ali, O., Chen, W., Feng, S., Wang, Q., L{\"u}, S., Long, M., Schneider, R., Sampathkumar, A., Godin, C., Traas, J., and Jiao, Y. (2020).
\newblock Microtubule-mediated wall anisotropy contributes to leaf blade flattening.
\newblock {\em Current Biology}, 30(20):3972--3985.

\bibitem[Zhu and Melrose, 2003]{zhu2003mechanics}
Zhu, H. and Melrose, J. (2003).
\newblock A mechanics model for the compression of plant and vegetative tissues.
\newblock {\em Journal of theoretical biology}, 221(1):89--101.

\end{thebibliography}

\end{document}